\documentclass[%
  aps,
  pre,
  superscriptaddress, letterpaper,
  longbibliography,
   reprint,
  showpacs,
  showkeys,
  amsfonts, amssymb, amsmath]{revtex4-1}

\usepackage{float} 
\usepackage{xcolor}
\usepackage[applemac]{inputenc}
\usepackage{graphicx,amssymb,amsmath}
\usepackage{epstopdf}							
\usepackage[english]{babel}

\begin{document}

\title{Impact of dissipation on the energy spectrum of experimental turbulence of gravity surface waves}
\author{Antoine Campagne, Roumaissa Hassaini, Ivan Redor, Jo\"el Sommeria, Thomas Valran, Samuel Viboud, Nicolas Mordant}
\email[]{nicolas.mordant@univ-grenoble-alpes.fr}
\affiliation{Laboratoire des Ecoulements G\'eophysiques et Industriels, Universit\'e Grenoble Alpes, CNRS, Grenoble-INP,  F-38000 Grenoble, France}

\begin{abstract}
We discuss the impact of dissipation on the development of the energy spectrum in wave turbulence of gravity surface waves with emphasis on the effect of surface contamination. We performed experiments in the Coriolis facility which is a 13-m diameter wave tank. We took care of cleaning surface contamination as well as possible considering that the surface of water exceeds 100~m$^2$. We observe that for the cleanest condition the frequency energy spectrum shows a power law decay extending up to the gravity capillary crossover (14 Hz) with a spectral exponent that is increasing with the forcing strength and decaying with surface contamination. Although slightly higher than reported previously in the literature, the exponent for the cleanest water remains significantly below the prediction from the Weak Turbulence Theory. By discussing length and time scales, we show that weak turbulence cannot be expected at frequencies above 3 Hz. We observe with a stereoscopic reconstruction technique that the increase with the forcing strength of energy spectrum beyond 3~Hz is mostly due to the formation and strenghtening of bound waves.
\end{abstract}

\maketitle

The effect of an oil film spread on the sea to calm the waves has been reported since Antiquity. This phenomenon is used in practice to detect remotely oil spills by radar probing the roughness of the sea surface~\cite{Wismann}. Experiments show that the maximum damping occurs usually for frequencies between 1 and 10~Hz (i.e. for wavelengths between 1~cm to 1~m)~\cite{Dorn,Cini,Alpers}. In the laboratory, the dedicated wave tanks are of typical size equal to a few times 10 m. In order to fit enough wavelengths in the tank to observe significant phenomena, the typical excitation of waves occurs most often at wavelengths of the order of 1~m (about 1~Hz for deep water waves) or slightly larger. In the wave turbulence framework, energy is expected to cascade in wavelength space from forcing scales to small dissipative scales~\cite{R1,R2,R3}. It means that the range of wavelengths over which the cascade occurs is precisely the one in which the damping by surface contamination is supposed to be the most efficient. This damping is most likely impacting significantly the nonlinear cascade and it maybe one of the reasons that explain the discrepancy between laboratory observations and theoretical predictions from the weak turbulence theory~\cite{R10,Deike}. Indeed considering the surface of wave tanks covering several hundreds square meter, it is very challenging to achieve a perfect control of the quality of the water surface, so that surface contamination is hard to avoid. {Dissipation is known to cause a steepening of turbulent wave elevation spectra as was reported for elastic waves in a thin plate~\cite{R23,Humbert} and for capillary-gravity waves~\cite{Deikedis,Deikedis2}.}

Here we report experiments dedicated to observe the impact of surface contamination on wave turbulence of surface gravity-capillary waves. We also discuss more generally the impact of dissipation of the development of the energy cascade due to wave turbulence of gravity surface waves. 

\section{Wave damping by molecular films at the surface of water}

\begin{figure}[!htb]
\includegraphics[width=9cm]{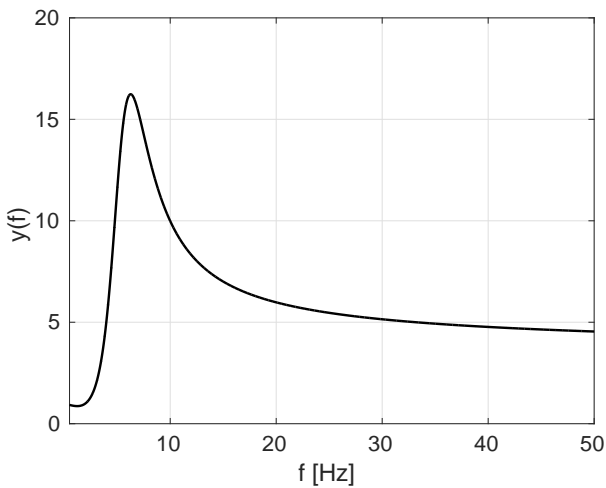}
\caption{Additional dissipation factor $y(f)$ as a function of frequency using Przadka {\it et al.} parameters in equation (\ref{y}) for a commercial paint pigment~\cite{Przadka2011}. A maximum is observed at about 4~Hz with a peak value close to 15.}
\label{yfig}
\end{figure}

The physical mechanism hidden behind this spectacular phenomenon is the modification of the tangential stress 
boundary condition at the air/water interface. For a perfectly clean interface, the tangential stress should vanish due to the much lower density of air. When a monomolecular film is present at the interface, the tangential stress does not vanish anymore due to concentration gradients resulting from the elongation or compression of the film. In the presence of a film, the boundary layer can sustain longitudinal waves (referred to as Marangoni waves) due to the viscoelastic properties of the film~\cite{Miles,Alpers, Henderson1}. The consequence of the modification of the boundary condition is the appearance of strong velocity gradients in the boundary layer which are maximum when the Marangoni waves are resonant with the gravity-capillary waves. This resonance leads to a much stronger dissipation that the one of a perfectly clean surface. Following Alpers \& H\"uhnerfuss~\cite{Alpers}, let us write the wave vector as $\kappa=k+i\Delta$. $\kappa$ is complex due to the attenuation ($k$ and $\Delta$ are real numbers). For a perfectly clean surface, the damping coefficient $\Delta_0$ is equal to
\begin{equation}
\Delta_0=\frac{4k^2\eta\omega}{\rho g+3\gamma k^2}\, ,
\label{delta0}
\end{equation}
where $\eta$ is the dynamic viscosity, $\rho$ is the density of the fluid, $\gamma$ is the surface tension of pure water, $g$ is the acceleration of gravity and $\omega$ is the angular temporal frequency. Alpers \& H\"uhnerfuss reports the calculation of the extra dissipation factor $y(\omega)=\frac{\Delta}{\Delta_0}$ due to the surface film:
\begin{eqnarray}
y&=&\frac{1+X(\cos\theta-\sin\theta)+XY-Y\sin\theta}
{1+2X(\cos\theta-\sin\theta)+2X^2}\label{y}\\
X&=&\frac{|E|k^2}{(2\rho\eta\omega^3)^{1/2}},\quad \quad Y=\frac{|E|k}{4\omega\eta}
\end{eqnarray}
where the complex dilational modulus of the film is $E=|E|e^{-i\theta}$ (see fig.~\ref{yfig} for an example). Alpers \& H\"uhnerfuss investigated the variation of $y$ for various contaminants and they observed typically that the peak of the resonance occurs at frequencies between 3 and 7 Hz and that the maximum overdamping reaches easily a factor 10 or even higher. This frequency range corresponds to wavelengths between 4 and 20 centimeters. At lower frequencies, for these surfactants, $y$ goes back to one at about 1~Hz (i.e. for metric wavelengths).  
The existence of the resonance was confirmed experimentally by Cini \& Lombardini~\cite{Cini}. The variation of interfacial surfactant concentration due to the interaction of surface waves with the surfactant layer was directly observed by Strickland {\it et al.}~\cite{Strickland}. {Note that dissipation occurs also through boundary layers at the bottom and at the vertical walls of the tank~\cite{Henderson2}. This contribution is expected to be significant for large wavelength and to be independent of the surface contamination.}

\section{The issue of spectra in experimental wave turbulence}

The weak turbulence theory (WTT) is aimed at describing the statistical properties of a wave assembly, notably in the out of equilibrium case in which waves are forced at relatively large scale and dissipated at the smallest scales~\cite{R9,R1,R2,R3}. It is based on the hypothesis of weak nonlinearity such that a scale separation exists between the period $T$ of the wave and $T_{NL}$ the time scale of the slow evolution of the wave energy due to the nonlinear coupling with the ensemble of all the other waves. $T_{NL}$ can be seen as the timescale of the correlation of the wave amplitude modulation (see \cite{Miquel3}). Under this hypothesis, a multiscale analysis can be developed that predicts the occurrence of an energy cascade. For waves propagating on a 2D surface, the energy cascade is direct i.e. it transfers energy to small scales. The WTT also provides estimates of the surface deformation spectrum $E^{\eta}(k)$ in either gravity or capillary regimes:
\begin{eqnarray}
E^{\eta}(k)&\propto& g^{1/2}P^{1/3}k^{-5/2}\quad \textrm{(gravity waves),}\\
E^{\eta}(k)&\propto& \gamma^{1/4}P^{1/2}k^{-7/4}\quad \textrm{(capillary waves),}
\end{eqnarray}
where $g$ is the gravity acceleration and $P$ is the energy flux in the cascade.
Using the dispersion relation
\begin{equation}
\omega_{LDR}^2=gk+\frac{\gamma}{\rho}k^3
\label{rd}
\end{equation}
at either small or large $k$ one can translate the $k$ spectra prediction into frequency spectra as
\begin{eqnarray}
E^{\eta}(\omega)\propto& gP^{1/3}\omega^{-4}\quad &\textrm{(gravity waves),}\\
E^{\eta}(\omega)\propto& \left(\frac{\gamma}{\rho}\right)^{1/6}P^{1/2}\omega^{-17/6}\quad &\textrm{(capillary waves).}
\end{eqnarray}

\begin{figure}[!htb]
\includegraphics[width=8.5cm]{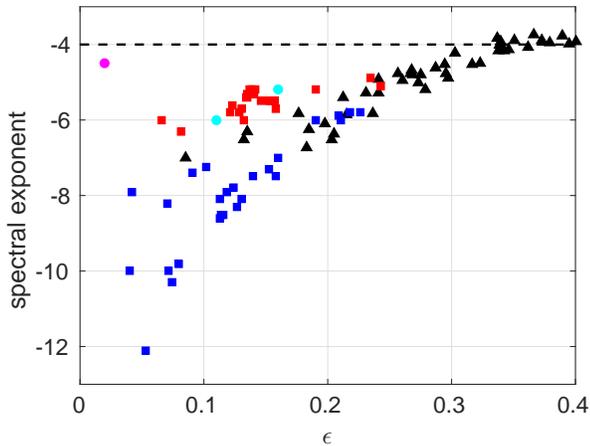}
\caption{Measured spectral exponent $\alpha$ of the temporal spectrum as function of the typical steepness of the waves $\epsilon$ (changed by tuning the magnitude of the forcing and dependent as well on the surface contamination). Dark triangles are previous measurements reported by Deike et al. \cite{Deike}. The blue and red squares are measurements by Aubourg~\cite{AubourgPhD,Aubourg2017} with and without filtration. The single purple point is an in-situ measurement of gravity waves in the Black sea \cite{Leckler}. Our present data for the cleanest case are the two cyan dots.}
\label{figexp}
\end{figure}

Concerning gravity waves, some field measurements seem compatible with the theoretical prediction \cite{Leckler,Hwang,Romero,Melville,Lenain} but the natural conditions may not be matched to the theoretical hypotheses notably that of stationarity and isotropy. Several experiments have been aimed at confronting these predictions to well controlled experiments in the laboratory notably in the last decade~\cite{Denissenko2007,Nazarenko2016,Deike,Aubourg2017}. The measured spectra ``look" turbulent i.e. they show a wide range of excited scales down the forcing scale but the spectral exponent fitted to the data are significantly lower than the theoretical prediction~\cite{Nazarenko2016,Deike,Aubourg2017} both for $E^{\eta}(k)$ and $E^{\eta}(\omega)$. An example is shown in fig.~\ref{figexp} which is a compilation of exponents from Aubourg et al.~\cite{Aubourg2017} and Deike et al.~\cite{Deike} as a function of the wave steepness $\epsilon$. $\epsilon$ is defined as $\epsilon=2k_p \sigma_\eta$ where $k_p$ is the wavenumber of the main peak of the spectrum and $\sigma_\eta$ is the standard deviation of the surface height. This definition is quite usual in oceanography and has been shown in \cite{AubourgPhD} to be indeed a good estimate of the slopes of the water surface. In the data from Aubourg~\cite{Aubourg2017}, two sets of exponents are visible. In red, the data correspond to a strongly contaminated water surface. The blue set of exponents corresponds to a case for which the surface was cleaned up to some extent by using a skimmer. The impact of surface contamination is very strong: for instance for a typical steepness of 0.1, the exponent is reduced by more than 2 between the filtered case and the cleaner case. Note that the distinction between the two sets reduces when the steepness of the wave (i.e. the nonlinearity) is increased. For very steep waves, the exponent reaches the theoretical value of $-4$ but in a strongly nonlinear regime at odds with the range of validity of the theoretical prediction for weak turbulence. { Nazarenko {\it et al.}~\cite{R10} report that it neither fits with the $k$ and $\omega$ spectra proposed by Philips~\cite{Phillips} or Kuznetsov~\cite{Kuz} for singular or overturning waves respectively.}

\section{Experiments in the Coriolis facility}

\subsection{Experimental setup}

\begin{figure}[!htb]
(a)\includegraphics[width=6cm]{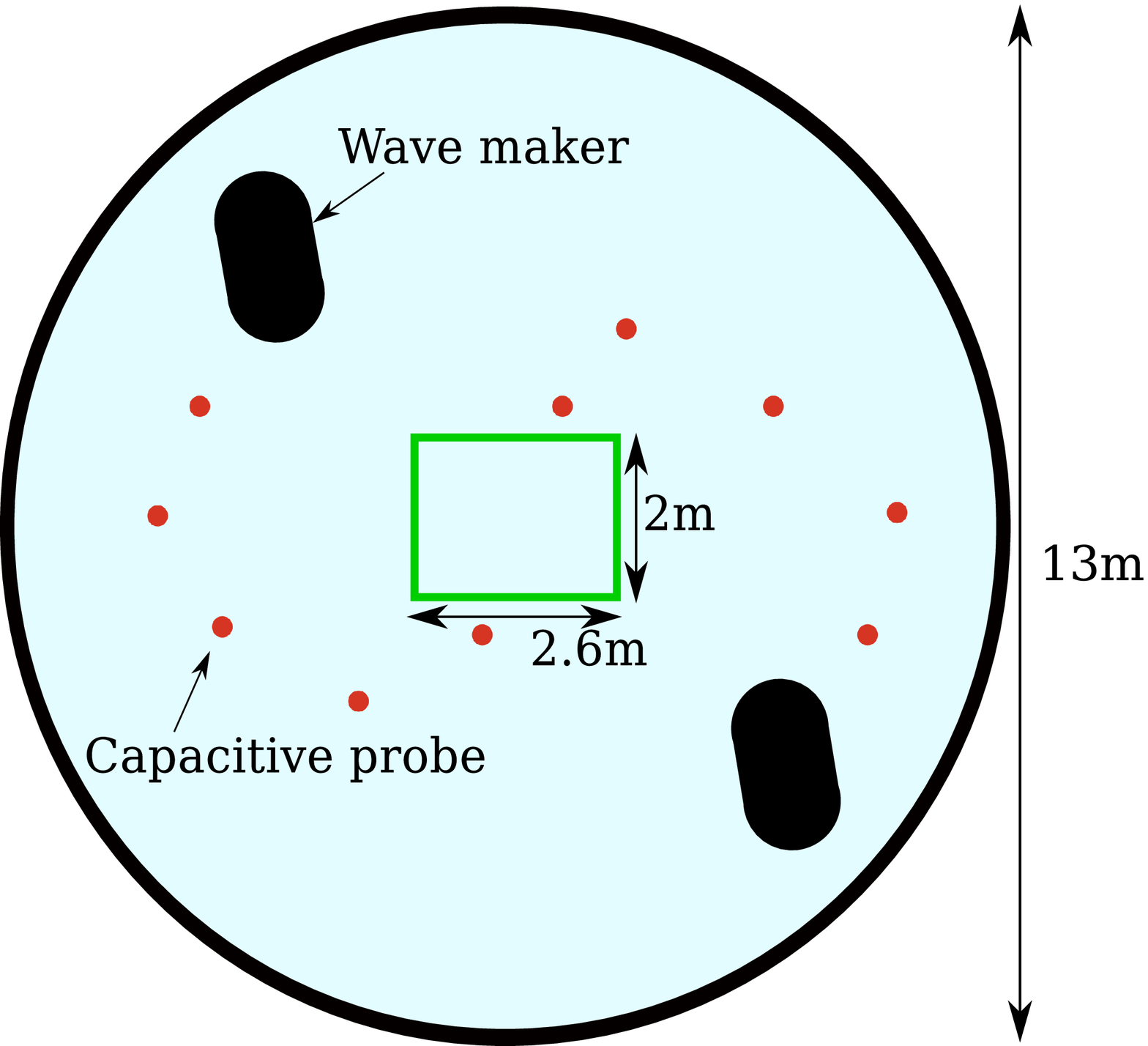}
(b)\includegraphics[width=6cm]{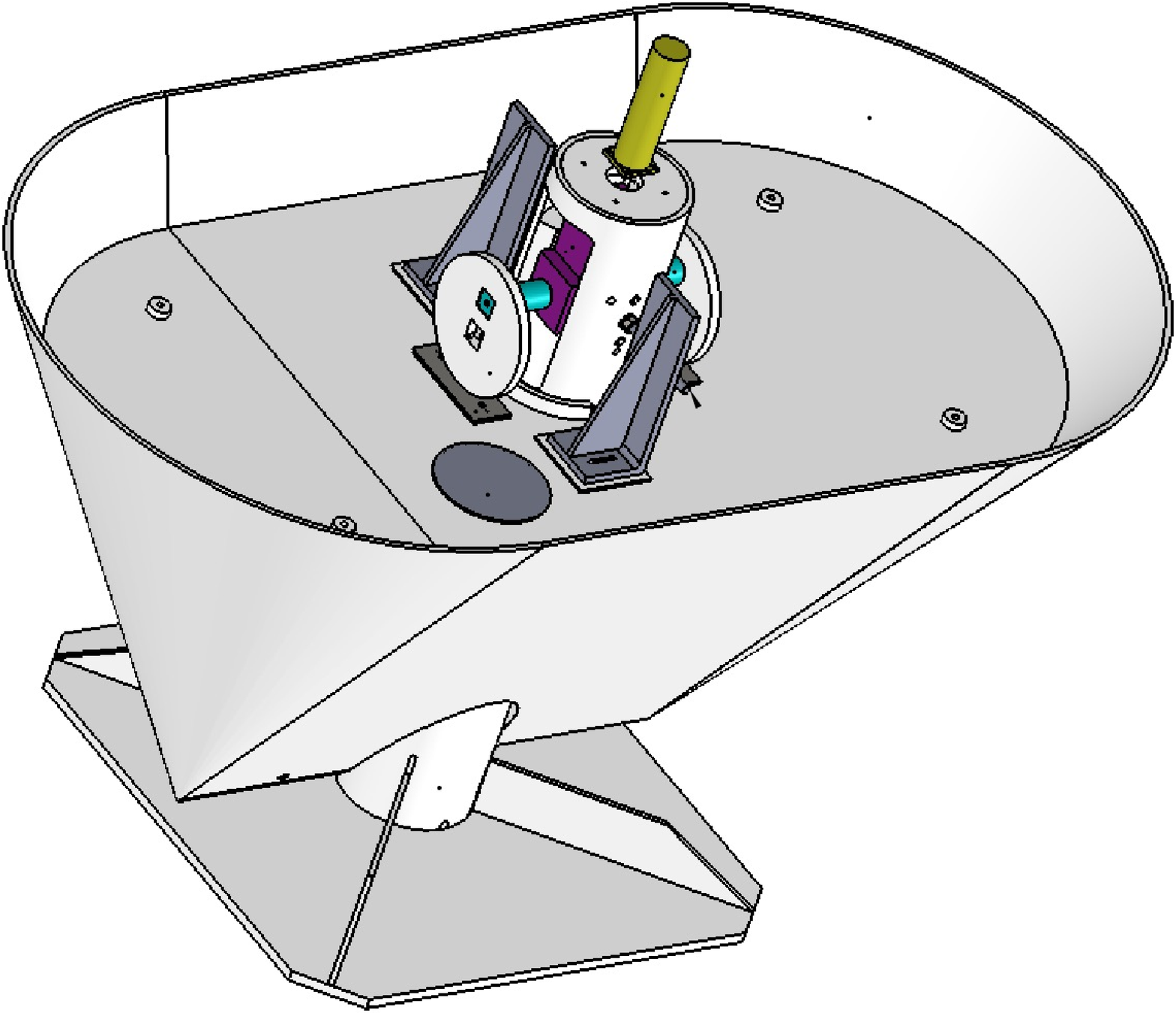}
\caption{(a) Global schematics of the setup in the Coriolis facility (seen from above). The tank is 13m in diameter and the water is 0.9~m deep. The position of the two wavemakers is shown (black ovals) and that of the 10 capacitive probes is shown as red dots. The field of view of the stereoscopic reconstruction (see section~\ref{skw}) is the green rectangle at the center. (b) schematics of a wavewaker. It is a wedge wavemaker (horizontal size 2m$\times$ 1m at the top) set into vertical oscillation by an eccentric cam. The off-center distance is 20~mm in the two experiments reported here. The rotation frequency is changed randomly in a given interval $0.585 \pm 0.15$~Hz and $0.78\pm 0.15$~Hz for the two experiments.}
\label{setup}
\end{figure}
The setup is very similar to that of Aubourg {\it et al.} (fig.~\ref{setup})\cite{AubourgPhD,Aubourg2017}. The wave tank is circular with a diameter equal to 13~m and the water depth is $h=0.9$~m. The water surface is maintained as clean a possible by pumping the surface through a skimmer located near the wall and by flowing the pumped water through an active carbon filter. The water is then reinjected near the wall at a location diametrically opposite to the skimmer. Wave are generated by two wedge wavemakers oscillating vertically at a randomly modulated frequency (fig.~\ref{setup}(b)). The wavemakers have been upgraded compared to Aubourg {\it et al.}. They keep their wedge geometry but with round ends so that to have a less directional generation. 
Two conditions of forcing were studied: a center frequency $f_0$ either equal to $0.585 \pm 0.15$~Hz (called ``weak'' case, steepness $\epsilon=0.11$) or $0.78\pm 0.15$~Hz (called ``strong'' case, $\epsilon=0.16$) with the same amplitude ($2$~cm) of vertical oscillation of the wavemaker. 
The filtration is not operating during the experiments so that it does not induce a spurious current in the tank. The surface elevation is recorded by using 10 capacitive wave gauges (their positions are shown in fig.~\ref{setup}). Surface tension is known to be very sensitive to surface contamination. Thus we measure surface tension by sampling the water surface regularly before and after the record once waves are damped. For reasons of convenience, the sample is taken at the periphery of the tank. The measured surface tension displays some variability due to sample collection and also due to a possible inhomogeneity of the surface contamination when waves are present. Indeed a weak but visible surface current is induced by the waves that can advect patches of surface films. Nevertheless our protocol allowed us to achieve a reasonable repeatability of the measurement. We estimate the confidence of the measurement close to $\pm 5mN/m$. After several days of filtration the measured surface tension was $74$~mN/m which was the value measured from fresh tap water at the same temperature. For clean water the repeatability of the measurement was actually significantly better ($\pm 2mN/m$). 

\subsection{Frequency spectra}
\begin{figure}[!htb]
\includegraphics[width=8.5cm]{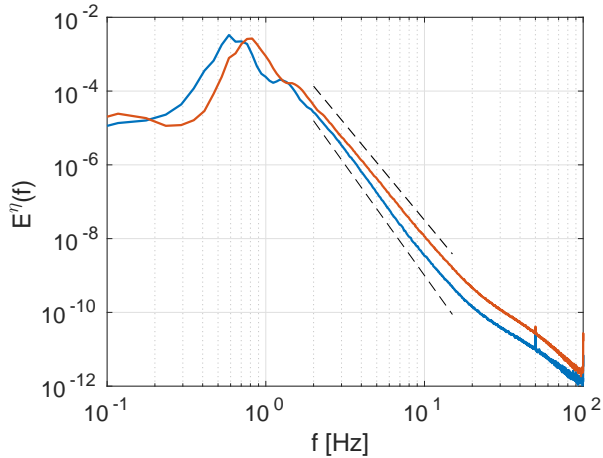}
\caption{Spectra after several days of almost continuous filtration. Top: ``strong'' case, bottom: ``weak'' case. The spectra have been averaged over several capacitive probes. The dashed lines correspond to a decay $1/f^{5.2}$ (top) and $1/f^6$ (bottom). The gravity-capillarity crossover occurs at 14Hz which corresponds to the observed change of slope of the spectrum.  The signal reaches the noise level at about 100~Hz.}
\label{figsp}
\end{figure}
Figure~\ref{figsp} shows typical spectra obtained after several days of almost continuous filtration of the surface. The weakest case displays a spectral exponent close to $-6$ and the strongest case an exponent close to $-5$. Although the duration of filtration was much longer than in Aubourg~{\it et al.} \cite{AubourgPhD,Aubourg2017} the spectral exponents do not exceed the upper limit of the previous data (fig.~\ref{figexp}).

\begin{figure}[!htb]
(a)\includegraphics[width=8cm]{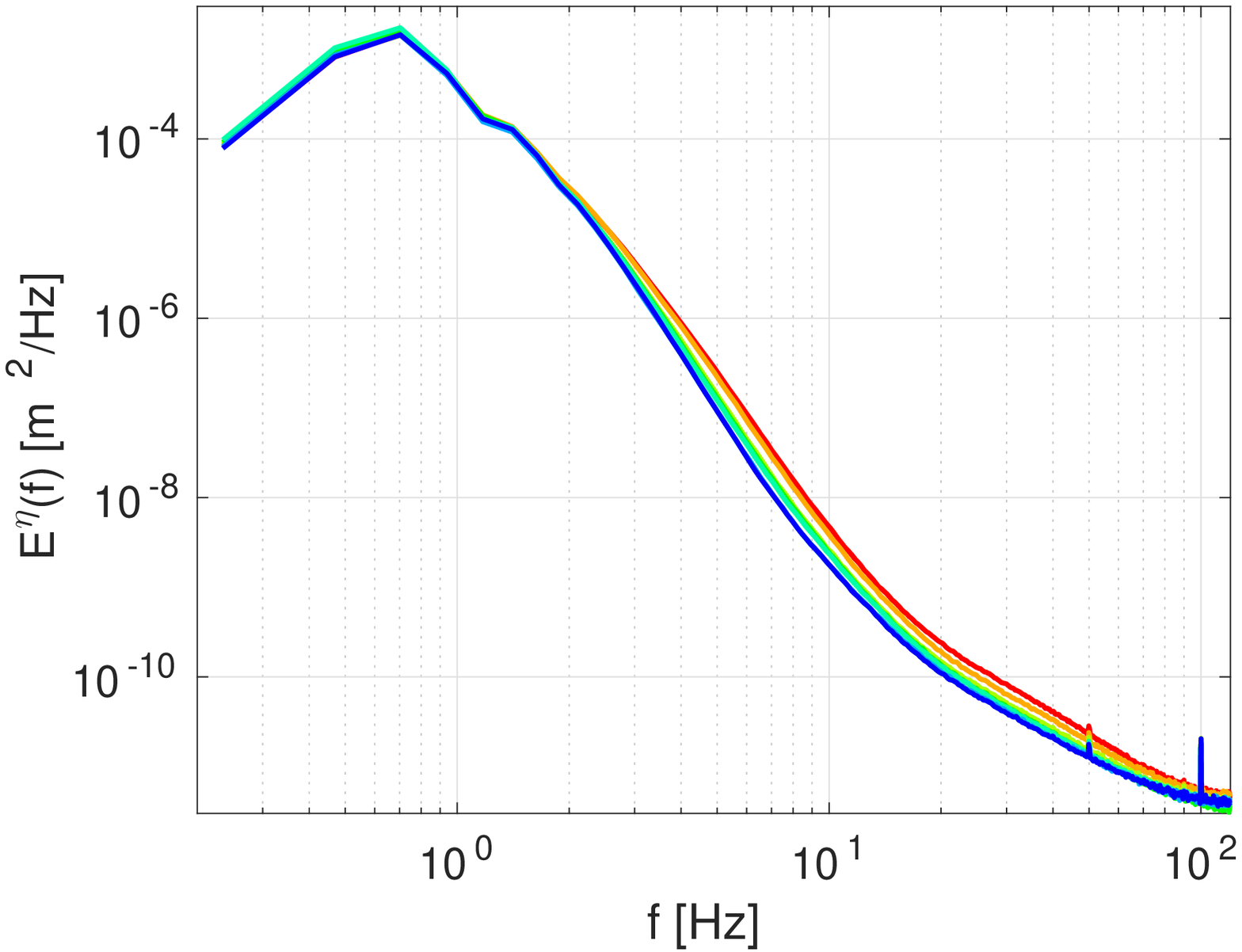}
(b)\includegraphics[width=8cm]{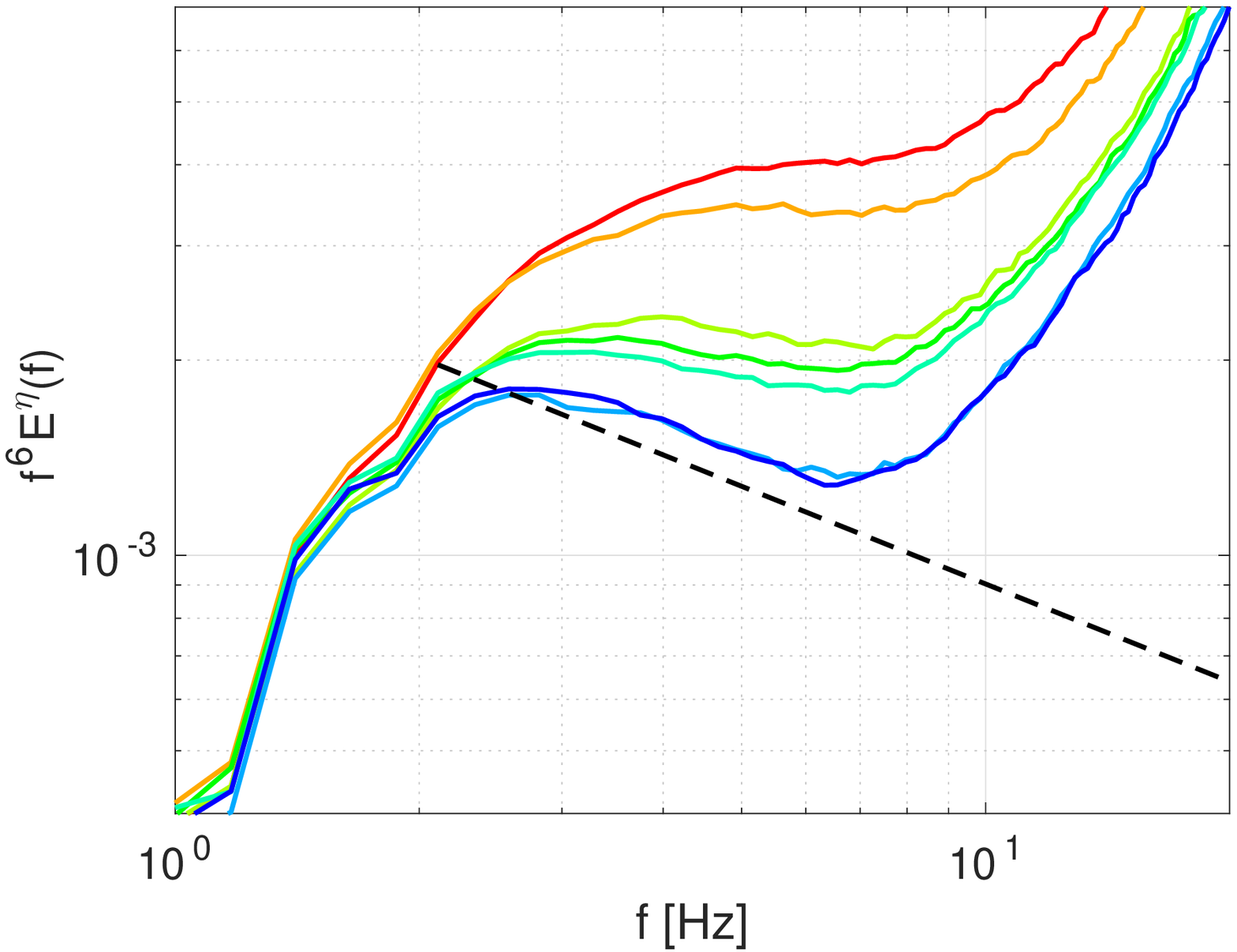}
\caption{Evolution of the wave spectrum versus time for weakest forcing. Each curve corresponds to an average over the 10 probes and a time average over one hour of continuous recording (7 successive records from red to blue). (a) Wave elevation spectrum $E^\eta(f)$. (b) compensated spectrum $f^6E^\eta(f)$ in the low frequency range. The dashed line is $\propto f^{-0.5}$.}
\label{spt}
\end{figure}
In a first experiment, after a long cleaning of the surface, we stop the filtration and record the wave elevation over several hours (each record being one hour long). The evolution of the spectra with time is shown in fig.~\ref{spt}. In fig.~\ref{spt}(b) the spectra have been compensated by $f^{6}$ in order to highlight the change of the slope. The spectra exponent is seen to decay continuously with time from $-6$ to $-6.5$. At the same time the surface tension decayed monotonously from 74~mN/m to 57 mN/m. The final value of the surface tension is typical of that observed for water exposed to air~\cite{Przadka2011}. The spectrum of the wave elevation is thus sensitive to the surface condition and is a good indicator of surface contamination for a given forcing condition.

\begin{figure}[!htb]
\includegraphics[width=8.5cm]{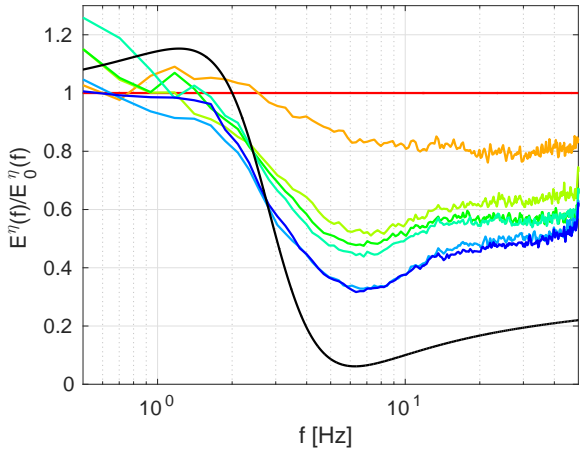}
\caption{Temporal evolution of the wave spectrum for weak forcing relative to the initial spectrum (same colors as Fig.~\ref{spt}). This representation highlights the deficit of energy in the spectrum as the surface becomes contaminated. The solid black curve represents the extra dissipation factor $1/y(f)$ for oleic acid (taken from Alpers et al.~\cite{Alpers}). For this dataset the forcing occurs at a central frequency of $0.585$~Hz. }
\label{figatt}
\end{figure}
In order to highlight the variation of attenuation as a function of frequency, we take the initial spectrum $E_0^\eta(f)$ as a reference and plot the ratio of the spectra divided by this initial value (Fig.~\ref{figatt}). The frequencies up to about 2~Hz are unaffected by the extra damping but higher frequencies are strongly diminished. A maximum damping efficiency is observed at about $6.5$~Hz that can reach $60\%$ for the latest spectra. The black  curve is the variation of $1/y(f)$ for oleic acid with parameters taken from Alpers {\it et al.} \cite{Alpers} that happens to have a maximum at the same frequency. This is mostly a coincidence as no oleic acid has been added to the water surface. The physico-chemical nature of the contamination is unknown as the contamination comes from dust falling on the surface and solvants from paint and plastic parts immersed in the water. Nevertheless the shape of $y(f)$ for oleic acid is quite typical and this comparison supports the fact that the increasing damping of our spectra comes from surface contamination.  The shape of the normalized spectra is actually qualitatively similar to that of $1/y(f)$ for oleic acid. Nevertheless the impact of surface damping is more complex than just filtering as it affects the nonlinear cascade that provides energy to waves. Damping at a given frequency must impact the waves at higher frequencies as the flux that feeds them must be reduced. Thus, in place of having a constant energy flux, the flux is progressively reduced as the frequency increases. 

\begin{figure}[!htb]
(a) \includegraphics[width=8cm]{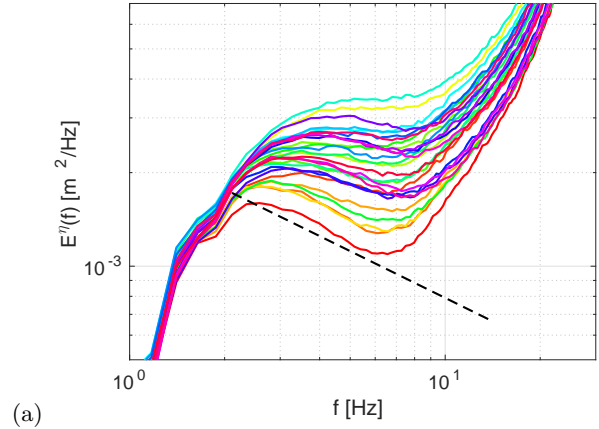}
(b) \includegraphics[width=8cm]{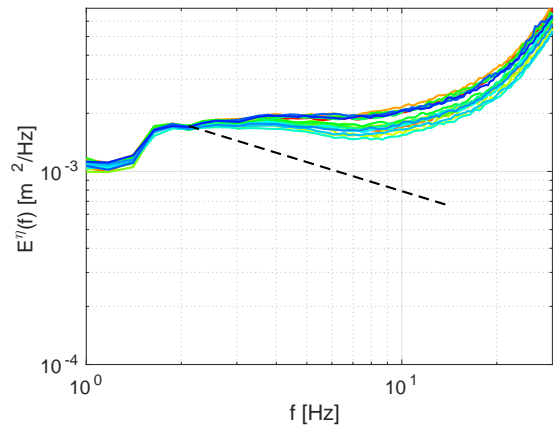}
\caption{Variability of spectra measured in similar conditions over time. (a) \& (b) weak and strong forcing respectively. In (a) we show 28 spectra (averaged over the 10 probes) corresponding each to a 1 hour long record. In (b) we show 19 similar spectra. The spectra have been compensated by $f^6$ and $f^5$ respectively. The dashed line shows a $f^{-1/2}$ decay. The variability is less at stronger forcing.}
\label{figsplong}
\end{figure}
Figure \ref{figsplong} shows a collection of spectra for either weak or strong forcing in a stationary configuration. Each curve corresponds to a one hour record of the surface elevation. The typical experimental sequence is 1 hour-long filtration, then two successive 1 hour-long records then filtration again. After a day of recording the filtration is then operated overnight and the sequence restarts. We see that a quite strong variability of the spectra of the weak case is seen in spite of our efforts to cleanup the surface. By contrast, the variability of the spectra is much reduced in the strong case (although the cleaning sequence is the same) which is much less sensitive to the surface condition. This is remisniscent of the dispersion of the exponents in fig.~\ref{figexp} which is much more pronounced for low steepness than for the strongest ones.

\section{Discussion of time scales}

The core of the Weak Turbulence Theory is the hypothesis of scale separation between the linear period of the wave $T$ and the nonlinear time scale $T_{NL}$ over which the non linearity operates. In the theory the dissipation is supposed to occur only at vanishing scales so that the dynamics is conservative over most scales leading for gravity waves to the theoretical prediction:
\begin{equation}
E(\omega)=C\frac{gP^{1/3}}{\omega^{4}}
\end{equation}
where $C$ is a constant, $g$ is the gravity acceleration and $P$ is the energy flux~\cite{R2}.

In actual physical systems, viscosity operates at any scale but is most efficient at the smallest scales. One may expect the predictions of the Weak Turbulence theory to be valid if an additional scale separation exists between $T_{NL}$ and the dissipative time scale $T_d$ so that
\begin{equation}
T\ll T_{NL}\ll T_d
\end{equation} 
at least in an inertial range of scales.

If one assumes that the dissipative time is much larger than the nonlinear time scale, the energy flux is progressively but only slightly depleted as the cascade proceeds to small scales. In that case the overall structure of the KZ spectrum should be preserved so that the energy spectrum may keep the form:
\begin{equation}
E(\omega)=C\frac{gP(\omega)^{1/3}}{\omega^{4}}
\label{spwt}
\end{equation}
where the energy flux $P(\omega)$ is slowly decaying with $\omega$. If the nonlinear coupling is very local in scale then the kinetic equation can be written in frequency space (see Ref.~\cite{R2} for instance) and for frequencies higher than that of the forcing as
\begin{equation}
\frac{\partial E(\omega)}{\partial t}=-\frac{\partial P}{\partial \omega}-2\gamma(\omega)E(\omega)\, ,
\end{equation}
with the inclusion of dissipation. $\nu=\frac{\eta}{\rho}$ is the kinematic viscosity and $\gamma(\omega)=2\nu k^2=2\nu\omega^4/g^2$ is the dissipation rate.
With the above shape of the energy spectrum, this equation can be rewritten in the stationary case as 
\begin{equation}
\frac{\partial P}{\partial \omega}=-4\nu CP^{1/3}/g
\end{equation}
The flux is thus obviously decaying with $\omega$. This equation can be integrated as 
\begin{equation}
P(\omega)=\left(P(\omega_f)^{2/3}-\frac{8\nu C}{3g}(\omega-\omega_f)\right)^{3/2}
\label{decflux}
\end{equation}
with $\omega_f$ being the frequency of the energy injection (supposed to be narrow band around $\omega_f$).
Using Eqs.~(\ref{spwt}) and (\ref{decflux}), the spectrum is then
\begin{equation}
E(\omega)=C\frac{g}{\omega^{4}}\left(P(\omega_f)^{2/3}-\frac{8\nu C}{3g}(\omega-\omega_f)\right)^{1/2}\, .
\label{corsp}
\end{equation}
The spectrum is thus expected to be slightly steeper than the KZ prediction in the limit $T_d\gg T_{NL}$. Unfortunately we cannot test directly this prediction as $P$ is unknown in our experiments as it is extremely difficult to measure it.

\begin{figure}[!htb]
(a)\includegraphics[width=8cm]{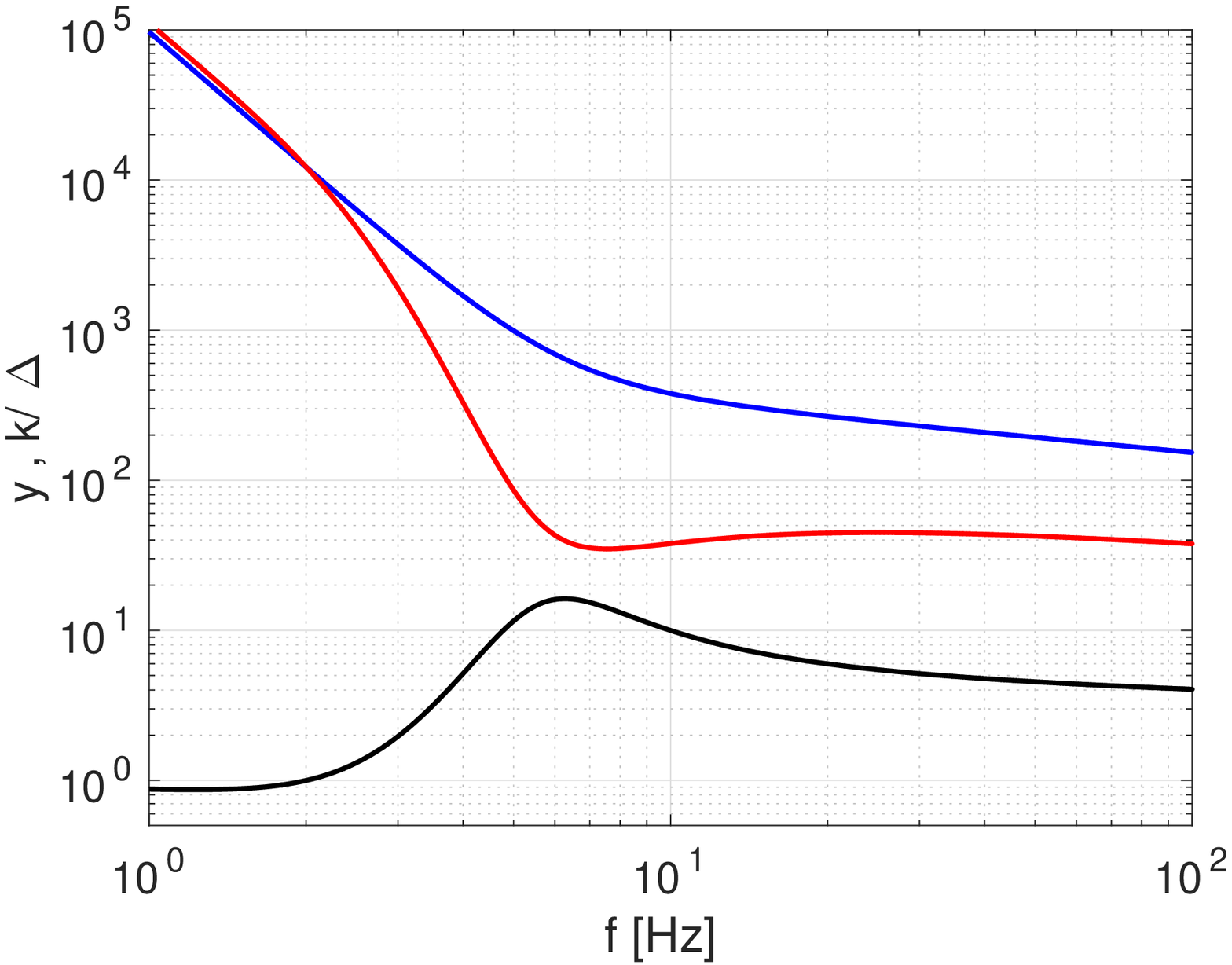}
(b)\includegraphics[width=8cm]{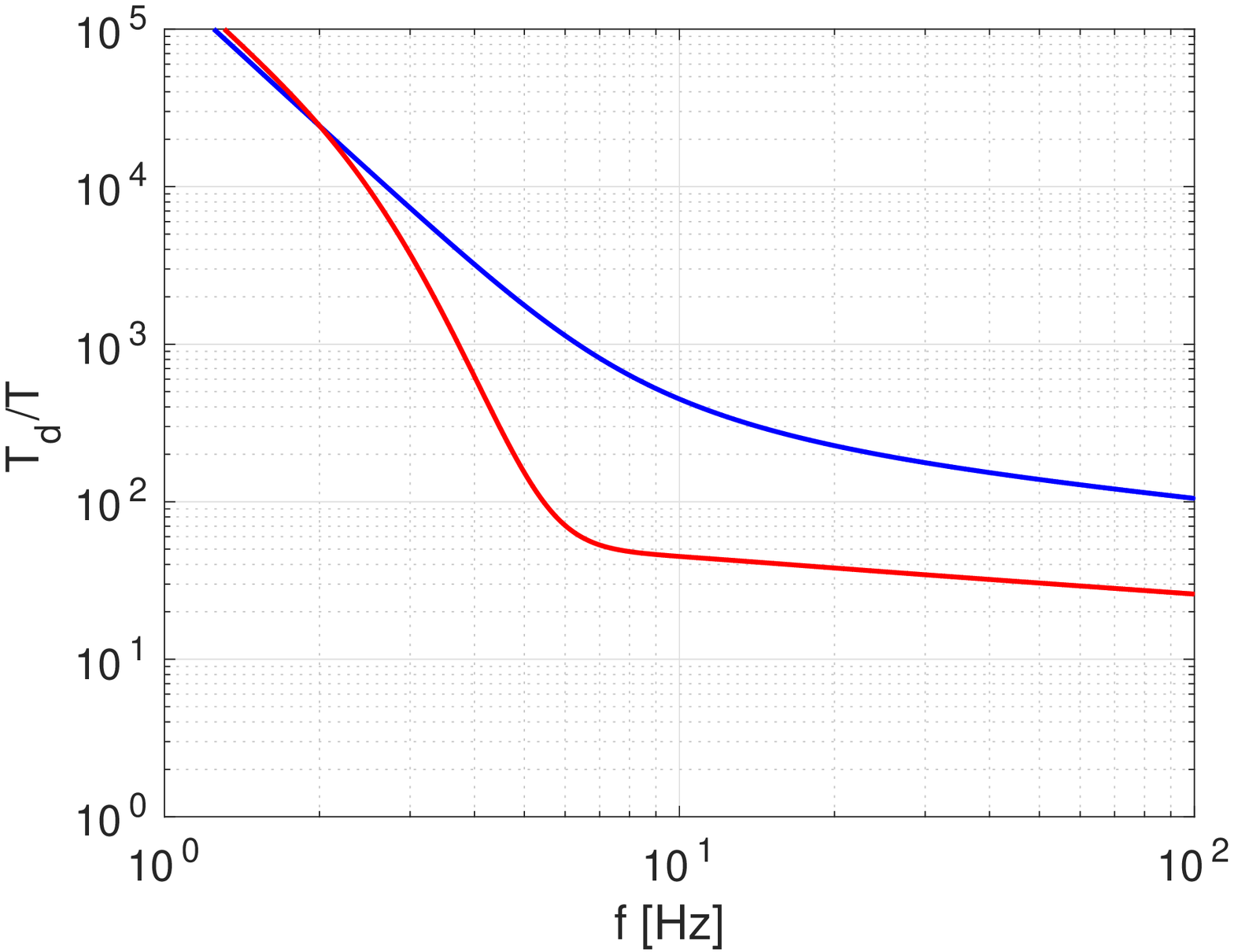}
\caption{(a)Ratio $\Delta/k$ which provides the dissipation length scale in terms of wavelengths. Blue: case of perfectly clean surface. Red: case of contaminated surface with oleic acid. Black: $y(f)$ for oleic acid (b) Timescale ratio of the dissipation scale $T_d$ over the period of the wave $T$. Same color code.}
\label{dislen}
\end{figure}
Figure~\ref{dislen}(a) shows the dissipative length scale expressed in terms of the wavelength i.e. the ratio $k/\Delta$. In the frequency interval $[1,10]$~Hz, for a perfectly clean surface, the ratio decays from $10^5$ wavelengths down to about 400. In terms of time scales, the imaginary part of the angular frequency is $2\nu k^2$ ($\Delta_0$ is actually computed by multiplying this value by the group velocity~\cite{Alpers}). Figure~\ref{dislen}(b) shows the ratio $T_d/T=\omega/2\nu k^2$. The variation of the time scale ratio takes naturally similar values to that of length scales. When the surface contamination is taken into account, the scale ratio is unchanged up to 2~Hz but may decay by more than one order of magnitude at the peak of $y(f)$ (the actual value depends on the pollutant) and falls below 100. 

{In the framework of the WTT}, due to the 4-wave interactions of gravity waves, the ratio $T_{NL}/T$ is expected to scale as $\epsilon^{-4}$ i.e. should be of order 10000 for $\epsilon=0.1$~\cite{Biven}. This can be achieved for frequencies below 2~Hz but not for higher frequencies even for a perfectly clean surface. For instance at 5~Hz ($\lambda\approx 7$~cm), the waves can propagate only over 1000 periods (with clear water) before being damped and over less than 100 periods for contaminated water. Thus one expects a very strong impact of dissipation over most frequencies even for very clean water and thus even the corrected spectrum (\ref{corsp}) should not be valid.

The ratio $T/T_{NL}$ actually depends on the frequency. In the gravity range {and in the kinetic regime}, if the wave spectrum follows the Kolmogorov-Zakharov spectrum, then following Newell {\it et al.} (\cite{Biven}, p.~544) one should have:
\begin{equation}
\frac{T}{T_{NL}}\approx \frac{1}{\omega n_k}\frac{\partial n_k}{\partial t}=CP^{2/3}k/g
\label{gnlt}
\end{equation}
where $n_k$ is the wave action spectrum. $\frac{\partial n_k}{\partial t}$ is estimated with the kinetic equation (see \cite{Biven} for details). The ratio $\frac{T}{T_{d}}$ should be in the gravity range 
\begin{equation}
\frac{T}{T_{d}}=\frac{4\pi\eta}{\rho g^{1/2}} k^{3/2}
\end{equation}
so that the ratio $T_{NL}/T_d$ is 
\begin{equation}
\frac{T_{NL}}{T_{d}}=\frac{4\pi\eta g^{1/2}}{C\rho P^{2/3}} k^{1/2}
\end{equation}

This ratio must remain much smaller than one but is increasing a $k^{1/2}$ so that at a given value of the nonlinearity the second inequality maybe broken at a critical wavenumber or frequency
\begin{eqnarray}
k_c&=&\left(\frac{C\rho P^{2/3}}{4\pi\eta g^{1/2}}\right)^2\\
\omega_c&=&\frac{C\rho g^{1/2} P^{2/3}}{4\pi\eta} \,\, .
\end{eqnarray}
$P^{2/3}$ is actually proportional to $\epsilon^4$ so the critical frequency is decaying very fast when $\epsilon$ is reduced ($k_c\propto \epsilon^8$, $\omega_c\propto \epsilon^4$) and can be very low for weak waves. Even for our cleanest conditions the spectral exponent is strongly reduced as compared to the KZ prediction. Furthermore it should be very sensitive to a slight contamination of the water surface that further reduces the dissipative time scale. These observations suggest that the frequencies of the waves in our experiment are beyond the critical frequency. The above scaling suggests that one should recover the KZ spectrum when increasing the nonlinearity. This appears consistent with the observation of spectral exponent becoming closer to the KZ prediction at larger steepness (Fig.~\ref{figexp}). The issue is then that the steepness may become so large that the hypothesis of small nonlinearity is broken and the waves are whitecapping. In this case one has $T\sim T_{NL}$ and the theory is not established in this limit.

At frequencies above the gravity-capillarity crossover the issue of time scale separation will be present as well. The pure capillary cascade is then a 3-wave process so that, {in the kinetic regime and for the KZ spectrum},
\begin{equation}
\frac{T}{T_{NL}}\approx \frac{1}{\omega n_k}\frac{\partial n_k}{\partial t}\propto P^{1/2}k^{-3/4}
\label{cnlt}
\end{equation}
and
\begin{equation}
\frac{T_{NL}}{T_{d}}\propto k^{5/4}P^{-1/2}
\end{equation}
Although the ratio $T/T_{NL}$ is getting smaller with $k$ (the cascade is getting less nonlinear as it proceeds), the ratio $T_{NL}/T_d$ is increasing faster with $k$. The critical wavenumber is thus
\begin{equation}
k_c\propto P^{2/5}
\end{equation}
$P^{1/2}$ is scaling as $\epsilon^2$ \cite{Biven} so that $k_c\propto \epsilon^{8/5}$. Thus, as for the case of gravity waves, $k_c$ is also decaying (although not as fast) when the nonlinearity is decreased.  

When the nonlinear time scale estimated from the kinetic equation (equations (\ref{gnlt}) \& (\ref{cnlt})) becomes comparable to $T_d$, the hypotheses underlying the computation of the collision term are not valid. Indeed, the obtention of the collision term involves taking a limit of large times that cannot be operated anymore if there is no scale separation. The frequency resonance condition results from the fact that only resonant waves can exchange a significant amount of energy, by a cumulative process, as $\epsilon$ goes to zero . Here we see that at weak enough a nonlinearity, the energy exchange can operate at most over a time $T_d$ which strongly reduces the efficiency of the energy transfers and thus should steepen the spectra. A distinct statistical theory must be developed. Based on the still existing scale separation $T/T_d\ll1$ a multiscale development may be relevant. For the kinetic theory to be valid one must have
$T\ll T_{NL}\ll T_d$
which is possible in laboratories only with very large flumes and most likely over a very narrow range of frequencies up to a few Hz (even with a clean water surface). Note that another limit exists which is the size of the wave tank which should be much larger than the wavelength which is not the case for meter wavelengths. We take advantage of reflexions on the walls to increase the effective propagation length at the expense of the presence of discrete modes (see \cite{Miquel3} and references therein for a discussion of finite size effects in another system supporting wave turbulence). 

Another observation in fig.~\ref{figatt} is that the attenuation is not as strong for frequencies higher than 10~Hz i.e. for capillary waves. If the cascade would be strongly local in wavenumber space, one would expect that the attenuation is a decaying function of the scale which is not what is observed. A non local nonlinear coupling mechanism has been reported by Aubourg \& Mordant~\cite{Aubourg2015} at the gravity-capillary crossover. This crossover occurs at $f=13.5$~Hz ($\lambda=1.7$~cm). This coupling has been observed to be quite strong because it involves only 3 waves. It is also nonlocal and couples short gravity waves ($1-2$~Hz) and capillary waves. The increase of the observed attenuation at frequencies larger that 7~Hz is most likely due to this mechanism that pumps energy directly from gravity waves. Note that being a 3-wave process, the nonlinear time of this mechanism is expected to be much shorter than that of the 4-wave process (scaling as $\epsilon ^2$ rather than $\epsilon^4$). Thus the 3-wave crossover coupling is likely to remain efficient even for contaminated water. Actually the next section reports space and time resolved measurements that show a very different explanation. 

{It must be noted also that another feature is operating in laboratory experiments which is the effect of the finite size of the flume. As mentioned in \cite{R10,Denissenko2007}, in finite basins, the linear modes are discrete. In order for a truly kinetic regime to develop (i.e. with continuous frequencies and wavenumbers) the nonlinear spectral widening must compensate for the mode separation. Thus the nonlinearity must be large enough, typically $\epsilon\geq 1/k_pL$ where $L$ is the size of the flume and $k_p$ is the value of the wavenumber at the peak of the spectrum. In our case the values are similar to that of refs.~\cite{R10,Denissenko2007} i.e. $\epsilon \geq 0.4$. We obviously do not reach such high values of the nonlinearity and this explains most likely that the spectral exponent does not reach the kinetic prediction as predicted in \cite{Nazarenko2013} even in the cleanest conditions.}

\section{$k$--$\omega$ spectrum}
\label{skw}

\begin{figure}[!htb]
\includegraphics[width=8.5cm]{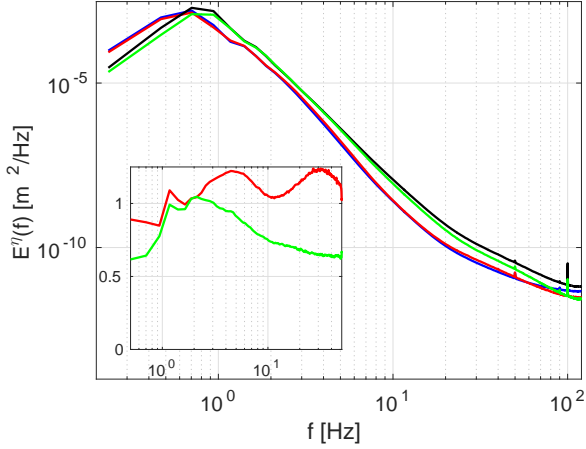}
\caption{Variation of spectra when floating particles are present or not. Blue/Red weak case, without/with particles resp. Black/Green strong forcing without/with particles resp. Insert: the spectra with particles have been normalized by the spectra without particles.}
\label{part}
\end{figure}

\begin{figure}[!htb]
(a)\includegraphics[width=7cm]{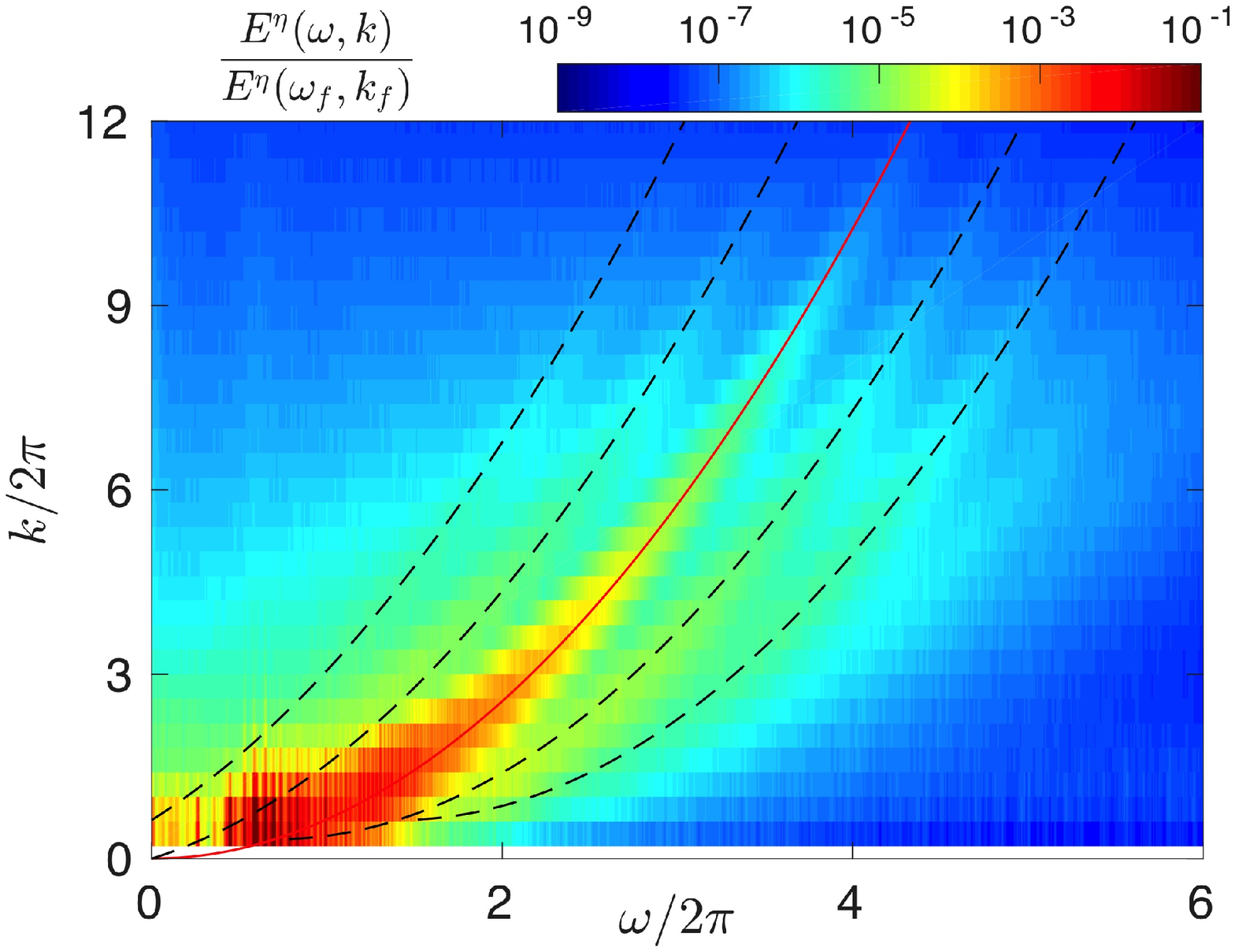}

(b)\includegraphics[width=7cm]{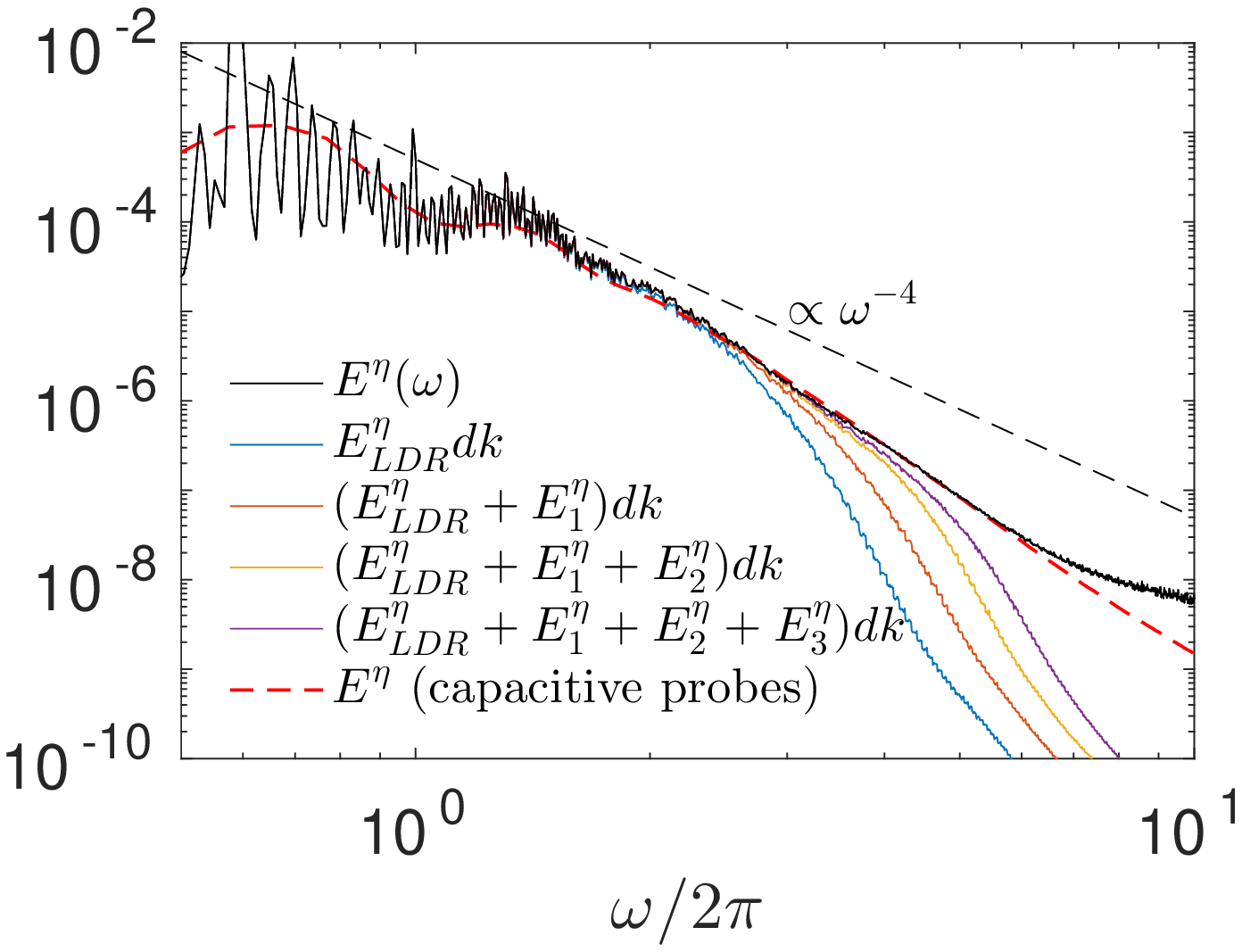}

(c)\includegraphics[width=7cm]{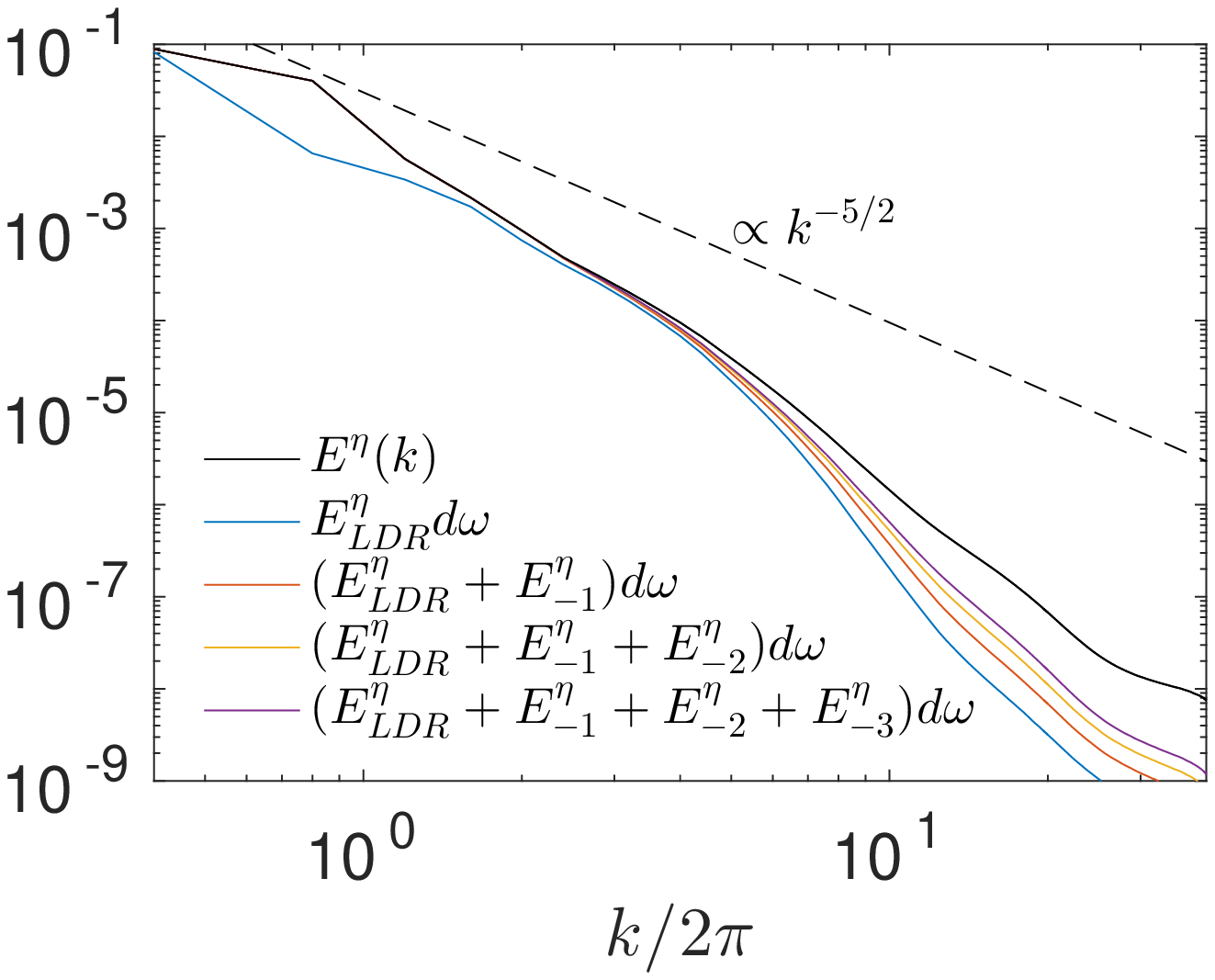}

\caption{(a) $k-\omega$ spectrum $E^\eta(k,\omega)$ for $\epsilon=0.11$ (``weak'' case). The spectrum has been normalized by its maximum value and is displayed in $\log_{10}$ scale. The red line is the linear dispersion relation (\ref{rd}). Dotted lines correspond to bound waves made from successive harmonics of the forcing peak and waves on the dispersion relation. (b) frequency spectrum constructed by adding successively the spectral contributions $E^{\eta}_n$ of the bound waves following $\omega^{(\pm n)}=\omega_{LDR}(k\mp nk_0)\pm n\omega_0$ with $n>0$ (see text). $E^\eta(\omega)$ is the full spectrum, $E^\eta_{LDR}(\omega)$ is the contribution on the linear dispersion relation and $E^\eta_n(\omega)$ is the contribution of the $n$-th bound waves as described in the text. The red dashed line is the spectrum obtained from the local probes. The divergence of the black and dashed red curves at about 7 Hz on the black line is due to the noise level of the stereoscopic measurement. (c) similar construction for the wavenumber spectrum $E^\eta(k)$.}
\label{kw}
\end{figure}

In order to check in more details the structure of the wave field, we use a stereoscopic technique to obtain a fully resolved (in space and time) measurement of the water surface (see \cite{AubourgPhD,Aubourg2017b} for information on the technique). We seed the surface of water with small (700~$\mu$m) buoyant particles to make the surface visible. The first question is the impact of the particles on the wave statistics. Strickland {\it et al.} \cite{Strickland} suggested that a mechanism similar to Marangoni damping could be operating when particles are floating at the surface. Figure~\ref{part} displays the comparison between surface elevation spectra obtained with clean water and with floating particles (at a surface concentration about 10 particles/cm$^2$). At the weakest forcing the spectrum seems actually slightly amplified. By contrast it is very weakly damped at the strongest forcing. This damping is weaker than that due to surface contamination and occurs at higher frequencies. Thus the mechanism of surface alteration due to floating particles seems quite distinct to that of chemical surface contamination. The weakness of the effect of adding the particles makes us confident that the particles do not alter the wave dynamics for scales much larger than the particle size.

The space-time spectrum $E^\eta(k,\omega)$ is shown in Fig.~\ref{kw}(a). The spectrum is obtained by performing a Fourier transform in both space and time (over time window of duration 125~s) providing $\eta(\mathbf k,\omega)$. The squared modulus of the Fourier transform is averaged over succesive time windows (Welch method) and integrated over directions of the wavevector $\mathbf k$ to give $E^\eta(k,\omega)$. The main energy component is a continuous line of energy on the dispersion relation that extends up to 4~Hz as expected for weak turbulence. Secondary energy lines are also visible on each side of the dispersion relation (highlighted by dashed lines) that correspond to so-called bound waves, which are not freely propagating waves following the quasilinear dispersion relation. They result from a triadic interaction between freely propagating waves. In our case, the observed lines can be obtained by assuming that the forcing peak at $(\mathbf k_f,\omega_f)$ is interacting with all free waves on the dispersion relation propagating in the same direction. The equation of the first line on the right of the dispersion relation in Fig.~\ref{kw}(a) is thus: $\omega^{(1)}=\omega_{LDR}(k-k_0)+\omega_0$ where $\omega(k)$ is the dispersion relation (\ref{rd}). The line on the left follows $\omega^{(-1)}=\omega_{LDR}(k+k_0)-\omega_0$. The lines further from the dispersion relation can be obtained assuming a similar interaction with successive harmonics of the forcing peak: $\omega^{(\pm n)}=\omega_{LDR}(k\mp nk_0)\pm n\omega_0$. It is worth noting that at frequencies higher than 4~Hz, almost all the energy lies in the bound waves. Fig.~\ref{kw}(b) shows the construction of the full frequency spectrum when adding progressively the energy lying on the bound waves (for $n>0$). It confirms that the contribution to the spectrum at frequencies higher than 4 Hz comes from those bound waves and not from an extension of the weakly non linear cascade to higher frequencies. The wave cascade seems to stop at 4~Hz in agreement with the above discussion on time scales. Fig.~\ref{kw}(c) shows a similar construction for the wavenumber spectrum when adding the bound waves (with $n<0$). It can be seen that the $k$-spectrum is much less sensitive to the bound waves than the frequency spectrum. This is due to the fact that the extension in $k$ of the bound waves is about the same than the main energy line lying on the linear dispersion relation and the energy of the bound waves remains smaller than that of the dispersion relation. 

In summary, the development of the weak energy cascade along the dispersion relation is strongly restricted by the viscous cutoff and even more restricted if the surface is contaminated. The extension of the frequency spectrum at higher frequencies observed for stronger forcing intensities is actually due to development of bound waves. In order to observe a weak turbulent cascade of gravity waves, one has to use much wider wave tanks in very large scale facilities in which a forcing could be achieved at lower frequency.

\begin{acknowledgments}
This project has received funding from the European Research Council (ERC) under the European Union's Horizon 2020 research and innovation programme (grant agreement No 647018-WATU).\end{acknowledgments}

\bibliography{bibliogravity}

\begin{thebibliography}{36}%
\makeatletter
\providecommand \@ifxundefined [1]{%
 \@ifx{#1\undefined}
}%
\providecommand \@ifnum [1]{%
 \ifnum #1\expandafter \@firstoftwo
 \else \expandafter \@secondoftwo
 \fi
}%
\providecommand \@ifx [1]{%
 \ifx #1\expandafter \@firstoftwo
 \else \expandafter \@secondoftwo
 \fi
}%
\providecommand \natexlab [1]{#1}%
\providecommand \enquote  [1]{``#1''}%
\providecommand \bibnamefont  [1]{#1}%
\providecommand \bibfnamefont [1]{#1}%
\providecommand \citenamefont [1]{#1}%
\providecommand \href@noop [0]{\@secondoftwo}%
\providecommand \href [0]{\begingroup \@sanitize@url \@href}%
\providecommand \@href[1]{\@@startlink{#1}\@@href}%
\providecommand \@@href[1]{\endgroup#1\@@endlink}%
\providecommand \@sanitize@url [0]{\catcode `\\12\catcode `\$12\catcode
  `\&12\catcode `\#12\catcode `\^12\catcode `\_12\catcode `\%12\relax}%
\providecommand \@@startlink[1]{}%
\providecommand \@@endlink[0]{}%
\providecommand \url  [0]{\begingroup\@sanitize@url \@url }%
\providecommand \@url [1]{\endgroup\@href {#1}{\urlprefix }}%
\providecommand \urlprefix  [0]{URL }%
\providecommand \Eprint [0]{\href }%
\providecommand \doibase [0]{http://dx.doi.org/}%
\providecommand \selectlanguage [0]{\@gobble}%
\providecommand \bibinfo  [0]{\@secondoftwo}%
\providecommand \bibfield  [0]{\@secondoftwo}%
\providecommand \translation [1]{[#1]}%
\providecommand \BibitemOpen [0]{}%
\providecommand \bibitemStop [0]{}%
\providecommand \bibitemNoStop [0]{.\EOS\space}%
\providecommand \EOS [0]{\spacefactor3000\relax}%
\providecommand \BibitemShut  [1]{\csname bibitem#1\endcsname}%
\let\auto@bib@innerbib\@empty
\bibitem [{\citenamefont {Wismann}\ \emph {et~al.}(1998)\citenamefont
  {Wismann}, \citenamefont {Gade}, \citenamefont {Alpers},\ and\ \citenamefont
  {Huhnerfuss}}]{Wismann}%
  \BibitemOpen
  \bibfield  {author} {\bibinfo {author} {\bibfnamefont {V.}~\bibnamefont
  {Wismann}}, \bibinfo {author} {\bibfnamefont {M.}~\bibnamefont {Gade}},
  \bibinfo {author} {\bibfnamefont {W.}~\bibnamefont {Alpers}}, \ and\ \bibinfo
  {author} {\bibfnamefont {H.}~\bibnamefont {Huhnerfuss}},\ }\bibfield  {title}
  {\enquote {\bibinfo {title} {{Radar signatures of marine mineral oil spills
  measured by an airborne multi-frequency radar}},}\ }\href@noop {} {\bibfield
  {journal} {\bibinfo  {journal} {International Journal of Remote Sensing}\
  }\textbf {\bibinfo {volume} {19}},\ \bibinfo {pages} {3607--3623} (\bibinfo
  {year} {1998})}\BibitemShut {NoStop}%
\bibitem [{\citenamefont {Dorn}(1966)}]{Dorn}%
  \BibitemOpen
  \bibfield  {author} {\bibinfo {author} {\bibfnamefont {W~G~Van}\ \bibnamefont
  {Dorn}},\ }\bibfield  {title} {\enquote {\bibinfo {title} {{Boundary
  dissipation of oscillatory waves}},}\ }\href@noop {} {\bibfield  {journal}
  {\bibinfo  {journal} {Journal Of Fluid Mechanics}\ }\textbf {\bibinfo
  {volume} {24}},\ \bibinfo {pages} {769} (\bibinfo {year} {1966})}\BibitemShut
  {NoStop}%
\bibitem [{\citenamefont {Cini}\ and\ \citenamefont {Lombardini}(1981)}]{Cini}%
  \BibitemOpen
  \bibfield  {author} {\bibinfo {author} {\bibfnamefont {R.}~\bibnamefont
  {Cini}}\ and\ \bibinfo {author} {\bibfnamefont {P.P.}\ \bibnamefont
  {Lombardini}},\ }\bibfield  {title} {\enquote {\bibinfo {title}
  {{Experimental evidence of a maximum in the frequency domain of the ratio of
  ripple attenuation in monolayered water to that in pure water}},}\
  }\href@noop {} {\bibfield  {journal} {\bibinfo  {journal} {Journal of Colloid
  and Interface Science}\ ,\ \bibinfo {pages} {125--131}} (\bibinfo {year}
  {1981})}\BibitemShut {NoStop}%
\bibitem [{\citenamefont {Alpers}\ and\ \citenamefont
  {Huhnerfuss}(1989)}]{Alpers}%
  \BibitemOpen
  \bibfield  {author} {\bibinfo {author} {\bibfnamefont {W.}~\bibnamefont
  {Alpers}}\ and\ \bibinfo {author} {\bibfnamefont {H.}~\bibnamefont
  {Huhnerfuss}},\ }\bibfield  {title} {\enquote {\bibinfo {title} {{The Damping
  of Ocean Waves by Surface Films: A New Look at an Old Problem}},}\
  }\href@noop {} {\bibfield  {journal} {\bibinfo  {journal} {Journal of
  Geophysical Research}\ }\textbf {\bibinfo {volume} {94}},\ \bibinfo {pages}
  {6251--6265} (\bibinfo {year} {1989})}\BibitemShut {NoStop}%
\bibitem [{\citenamefont {Zakharov}\ \emph {et~al.}(1992)\citenamefont
  {Zakharov}, \citenamefont {L'vov},\ and\ \citenamefont {Falkovich}}]{R1}%
  \BibitemOpen
  \bibfield  {author} {\bibinfo {author} {\bibfnamefont {V.~E.}\ \bibnamefont
  {Zakharov}}, \bibinfo {author} {\bibfnamefont {V.~S.}\ \bibnamefont {L'vov}},
  \ and\ \bibinfo {author} {\bibfnamefont {G.}~\bibnamefont {Falkovich}},\
  }\href@noop {} {\emph {\bibinfo {title} {Kolmogorov Spectra of Turbulence}}}\
  (\bibinfo  {publisher} {Springer},\ \bibinfo {address} {Berlin},\ \bibinfo
  {year} {1992})\BibitemShut {NoStop}%
\bibitem [{\citenamefont {Nazarenko}(2011)}]{R2}%
  \BibitemOpen
  \bibfield  {author} {\bibinfo {author} {\bibfnamefont {S.}~\bibnamefont
  {Nazarenko}},\ }\href@noop {} {\emph {\bibinfo {title} {Wave Turbulence}}}\
  (\bibinfo  {publisher} {Springer},\ \bibinfo {address} {Berlin},\ \bibinfo
  {year} {2011})\BibitemShut {NoStop}%
\bibitem [{\citenamefont {Newell}\ and\ \citenamefont {Rumpf}(2011)}]{R3}%
  \BibitemOpen
  \bibfield  {author} {\bibinfo {author} {\bibfnamefont {A.~C.}\ \bibnamefont
  {Newell}}\ and\ \bibinfo {author} {\bibfnamefont {B.}~\bibnamefont {Rumpf}},\
  }\bibfield  {title} {\enquote {\bibinfo {title} {Wave turbulence},}\
  }\href@noop {} {\bibfield  {journal} {\bibinfo  {journal} {Ann. Rev. Fluid
  Mech.}\ }\textbf {\bibinfo {volume} {43}} (\bibinfo {year}
  {2011})}\BibitemShut {NoStop}%
\bibitem [{\citenamefont {Nazarenko}\ \emph {et~al.}(2009)\citenamefont
  {Nazarenko}, \citenamefont {Lukaschuk}, \citenamefont {McLelland},\ and\
  \citenamefont {Denissenko}}]{R10}%
  \BibitemOpen
  \bibfield  {author} {\bibinfo {author} {\bibfnamefont {S.}~\bibnamefont
  {Nazarenko}}, \bibinfo {author} {\bibfnamefont {S.}~\bibnamefont
  {Lukaschuk}}, \bibinfo {author} {\bibfnamefont {S.}~\bibnamefont
  {McLelland}}, \ and\ \bibinfo {author} {\bibfnamefont {P.}~\bibnamefont
  {Denissenko}},\ }\bibfield  {title} {\enquote {\bibinfo {title} {Statistics
  of surface gravity wave turbulence in the space and time domains},}\
  }\href@noop {} {\bibfield  {journal} {\bibinfo  {journal} {J. Fluid Mech.}\
  }\textbf {\bibinfo {volume} {642}},\ \bibinfo {pages} {395} (\bibinfo {year}
  {2009})}\BibitemShut {NoStop}%
\bibitem [{\citenamefont {Deike}\ \emph {et~al.}(2015)\citenamefont {Deike},
  \citenamefont {Miquel}, \citenamefont {Gutierrez}, \citenamefont {Jamin},
  \citenamefont {Semin}, \citenamefont {Berhanu}, \citenamefont {Falcon},\ and\
  \citenamefont {Bonnefoy}}]{Deike}%
  \BibitemOpen
  \bibfield  {author} {\bibinfo {author} {\bibfnamefont {L.}~\bibnamefont
  {Deike}}, \bibinfo {author} {\bibfnamefont {B.}~\bibnamefont {Miquel}},
  \bibinfo {author} {\bibfnamefont {P.}~\bibnamefont {Gutierrez}}, \bibinfo
  {author} {\bibfnamefont {T.}~\bibnamefont {Jamin}}, \bibinfo {author}
  {\bibfnamefont {B.}~\bibnamefont {Semin}}, \bibinfo {author} {\bibfnamefont
  {M.}~\bibnamefont {Berhanu}}, \bibinfo {author} {\bibfnamefont
  {E.}~\bibnamefont {Falcon}}, \ and\ \bibinfo {author} {\bibfnamefont
  {F.}~\bibnamefont {Bonnefoy}},\ }\bibfield  {title} {\enquote {\bibinfo
  {title} {Role of the basin boundary conditions in gravity wave turbulence},}\
  }\href {\doibase 10.1017/jfm.2015.494} {\bibfield  {journal} {\bibinfo
  {journal} {J. Fluid Mech.}\ }\textbf {\bibinfo {volume} {781}},\ \bibinfo
  {pages} {196--225} (\bibinfo {year} {2015})}\BibitemShut {NoStop}%
\bibitem [{\citenamefont {Miquel}\ \emph {et~al.}(2014)\citenamefont {Miquel},
  \citenamefont {Alexakis},\ and\ \citenamefont {Mordant}}]{R23}%
  \BibitemOpen
  \bibfield  {author} {\bibinfo {author} {\bibfnamefont {B.}~\bibnamefont
  {Miquel}}, \bibinfo {author} {\bibfnamefont {A.}~\bibnamefont {Alexakis}}, \
  and\ \bibinfo {author} {\bibfnamefont {N.}~\bibnamefont {Mordant}},\
  }\bibfield  {title} {\enquote {\bibinfo {title} {Role of dissipation in
  flexural wave turbulence: from experimental spectrum to kolmogorov-zakharov
  spectrum},}\ }\href@noop {} {\bibfield  {journal} {\bibinfo  {journal} {Phys.
  Rev. E}\ }\textbf {\bibinfo {volume} {89}},\ \bibinfo {pages} {062925}
  (\bibinfo {year} {2014})}\BibitemShut {NoStop}%
\bibitem [{\citenamefont {Humbert}\ \emph {et~al.}(2013)\citenamefont
  {Humbert}, \citenamefont {Cadot}, \citenamefont {D\"uring}, \citenamefont
  {Josserand}, \citenamefont {Rica},\ and\ \citenamefont {Touz\'e}}]{Humbert}%
  \BibitemOpen
  \bibfield  {author} {\bibinfo {author} {\bibfnamefont {T.}~\bibnamefont
  {Humbert}}, \bibinfo {author} {\bibfnamefont {O.}~\bibnamefont {Cadot}},
  \bibinfo {author} {\bibfnamefont {G.}~\bibnamefont {D\"uring}}, \bibinfo
  {author} {\bibfnamefont {C.}~\bibnamefont {Josserand}}, \bibinfo {author}
  {\bibfnamefont {S.}~\bibnamefont {Rica}}, \ and\ \bibinfo {author}
  {\bibfnamefont {C.}~\bibnamefont {Touz\'e}},\ }\bibfield  {title} {\enquote
  {\bibinfo {title} {Wave turbulence in vibrating plates : the effect of
  damping},}\ }\href@noop {} {\bibfield  {journal} {\bibinfo  {journal} {EPL}\
  }\textbf {\bibinfo {volume} {102}},\ \bibinfo {pages} {30002} (\bibinfo
  {year} {2013})}\BibitemShut {NoStop}%
\bibitem [{\citenamefont {Deike}\ \emph {et~al.}(2012)\citenamefont {Deike},
  \citenamefont {Berhanu},\ and\ \citenamefont {Falcon}}]{Deikedis}%
  \BibitemOpen
  \bibfield  {author} {\bibinfo {author} {\bibfnamefont {L.}~\bibnamefont
  {Deike}}, \bibinfo {author} {\bibfnamefont {M.}~\bibnamefont {Berhanu}}, \
  and\ \bibinfo {author} {\bibfnamefont {E.}~\bibnamefont {Falcon}},\
  }\bibfield  {title} {\enquote {\bibinfo {title} {Decay of capillary wave
  turbulence},}\ }\href@noop {} {\bibfield  {journal} {\bibinfo  {journal}
  {Phys. Rev. E}\ }\textbf {\bibinfo {volume} {85}},\ \bibinfo {pages} {066311}
  (\bibinfo {year} {2012})}\BibitemShut {NoStop}%
\bibitem [{\citenamefont {Deike}\ \emph {et~al.}(2014)\citenamefont {Deike},
  \citenamefont {Berhanu},\ and\ \citenamefont {Falcon}}]{Deikedis2}%
  \BibitemOpen
  \bibfield  {author} {\bibinfo {author} {\bibfnamefont {L.}~\bibnamefont
  {Deike}}, \bibinfo {author} {\bibfnamefont {M.}~\bibnamefont {Berhanu}}, \
  and\ \bibinfo {author} {\bibfnamefont {E.}~\bibnamefont {Falcon}},\
  }\bibfield  {title} {\enquote {\bibinfo {title} {Energy flux measurement from
  the dissipated energy in capillary wave turbulence},}\ }\href@noop {}
  {\bibfield  {journal} {\bibinfo  {journal} {Phys. Rev. E}\ }\textbf {\bibinfo
  {volume} {89}},\ \bibinfo {pages} {023003} (\bibinfo {year}
  {2014})}\BibitemShut {NoStop}%
\bibitem [{\citenamefont {Miles}(1967)}]{Miles}%
  \BibitemOpen
  \bibfield  {author} {\bibinfo {author} {\bibfnamefont {JW}~\bibnamefont
  {Miles}},\ }\bibfield  {title} {\enquote {\bibinfo {title} {{Surface-wave
  damping in closed basins}},}\ }\href@noop {} {\bibfield  {journal} {\bibinfo
  {journal} {Proc. Roy. Soc. A}\ }\textbf {\bibinfo {volume} {297}},\ \bibinfo
  {pages} {459} (\bibinfo {year} {1967})}\BibitemShut {NoStop}%
\bibitem [{\citenamefont {Henderson}\ and\ \citenamefont
  {Segur}(2013)}]{Henderson1}%
  \BibitemOpen
  \bibfield  {author} {\bibinfo {author} {\bibfnamefont {D~M}\ \bibnamefont
  {Henderson}}\ and\ \bibinfo {author} {\bibfnamefont {H}~\bibnamefont
  {Segur}},\ }\bibfield  {title} {\enquote {\bibinfo {title} {{The role of
  dissipation in the evolution of ocean swell}},}\ }\href@noop {} {\bibfield
  {journal} {\bibinfo  {journal} {Journal Of Geophysical Research-Oceans}\
  }\textbf {\bibinfo {volume} {118}},\ \bibinfo {pages} {5074--5091} (\bibinfo
  {year} {2013})}\BibitemShut {NoStop}%
\bibitem [{\citenamefont {Przadka}\ \emph {et~al.}(2011)\citenamefont
  {Przadka}, \citenamefont {Cabane}, \citenamefont {Pagneux}, \citenamefont
  {Maurel},\ and\ \citenamefont {Petitjeans}}]{Przadka2011}%
  \BibitemOpen
  \bibfield  {author} {\bibinfo {author} {\bibfnamefont {A.}~\bibnamefont
  {Przadka}}, \bibinfo {author} {\bibfnamefont {B.}~\bibnamefont {Cabane}},
  \bibinfo {author} {\bibfnamefont {V.}~\bibnamefont {Pagneux}}, \bibinfo
  {author} {\bibfnamefont {A.}~\bibnamefont {Maurel}}, \ and\ \bibinfo {author}
  {\bibfnamefont {P.}~\bibnamefont {Petitjeans}},\ }\bibfield  {title}
  {\enquote {\bibinfo {title} {Fourier transform profilometry for water waves:
  how to achieve clean water attenuation with diffusive reflection at the water
  surface?}}\ }\href {\doibase 10.1007/s00348-011-1240-x} {\bibfield  {journal}
  {\bibinfo  {journal} {Exp. Fluids}\ }\textbf {\bibinfo {volume} {52}},\
  \bibinfo {pages} {519--527} (\bibinfo {year} {2011})}\BibitemShut {NoStop}%
\bibitem [{\citenamefont {Strickland}\ and\ \citenamefont
  {Daniels}(2015)}]{Strickland}%
  \BibitemOpen
  \bibfield  {author} {\bibinfo {author} {\bibfnamefont {M.}~\bibnamefont
  {Strickland}, \bibfnamefont {S.L.and~Shearer}}\ and\ \bibinfo {author}
  {\bibfnamefont {K.E.}\ \bibnamefont {Daniels}},\ }\bibfield  {title}
  {\enquote {\bibinfo {title} {{Spatiotemporal measurement of surfactant
  distribution on gravity--capillary waves}},}\ }\href@noop {} {\bibfield
  {journal} {\bibinfo  {journal} {Journal Of Fluid Mechanics}\ }\textbf
  {\bibinfo {volume} {777}},\ \bibinfo {pages} {523--543} (\bibinfo {year}
  {2015})}\BibitemShut {NoStop}%
\bibitem [{\citenamefont {Henderson}\ and\ \citenamefont
  {Miles}(1990)}]{Henderson2}%
  \BibitemOpen
  \bibfield  {author} {\bibinfo {author} {\bibfnamefont {D~M}\ \bibnamefont
  {Henderson}}\ and\ \bibinfo {author} {\bibfnamefont {John~W}\ \bibnamefont
  {Miles}},\ }\bibfield  {title} {\enquote {\bibinfo {title} {{Single-mode
  Faraday waves in small cylinders}},}\ }\href@noop {} {\bibfield  {journal}
  {\bibinfo  {journal} {Journal Of Fluid Mechanics}\ }\textbf {\bibinfo
  {volume} {213}},\ \bibinfo {pages} {95} (\bibinfo {year} {1990})}\BibitemShut
  {NoStop}%
\bibitem [{\citenamefont {Hasselmann}(1962)}]{R9}%
  \BibitemOpen
  \bibfield  {author} {\bibinfo {author} {\bibfnamefont {K.}~\bibnamefont
  {Hasselmann}},\ }\bibfield  {title} {\enquote {\bibinfo {title} {On the
  non-linear energy transfer in gravity-wave spectrum. part 1. general
  theory},}\ }\href@noop {} {\bibfield  {journal} {\bibinfo  {journal} {J.
  Fluid Mech.}\ }\textbf {\bibinfo {volume} {12}},\ \bibinfo {pages} {481--500}
  (\bibinfo {year} {1962})}\BibitemShut {NoStop}%
\bibitem [{\citenamefont {Miquel}\ and\ \citenamefont
  {Mordant}(2011)}]{Miquel3}%
  \BibitemOpen
  \bibfield  {author} {\bibinfo {author} {\bibfnamefont {B.}~\bibnamefont
  {Miquel}}\ and\ \bibinfo {author} {\bibfnamefont {N.}~\bibnamefont
  {Mordant}},\ }\bibfield  {title} {\enquote {\bibinfo {title} {Non linear
  dynamics of flexural wave turbulence},}\ }\href@noop {} {\bibfield  {journal}
  {\bibinfo  {journal} {Phys. Rev. E}\ }\textbf {\bibinfo {volume} {84}},\
  \bibinfo {pages} {066607} (\bibinfo {year} {2011})}\BibitemShut {NoStop}%
\bibitem [{\citenamefont {Leckler}\ \emph {et~al.}(2015)\citenamefont
  {Leckler}, \citenamefont {Ardhuin}, \citenamefont {Peureux}, \citenamefont
  {Benetazzo}, \citenamefont {Bergamasco},\ and\ \citenamefont
  {Dulov}}]{Leckler}%
  \BibitemOpen
  \bibfield  {author} {\bibinfo {author} {\bibfnamefont {F.}~\bibnamefont
  {Leckler}}, \bibinfo {author} {\bibfnamefont {F.}~\bibnamefont {Ardhuin}},
  \bibinfo {author} {\bibfnamefont {C.}~\bibnamefont {Peureux}}, \bibinfo
  {author} {\bibfnamefont {A.}~\bibnamefont {Benetazzo}}, \bibinfo {author}
  {\bibfnamefont {F.}~\bibnamefont {Bergamasco}}, \ and\ \bibinfo {author}
  {\bibfnamefont {V.}~\bibnamefont {Dulov}},\ }\bibfield  {title} {\enquote
  {\bibinfo {title} {{Analysis and Interpretation of Frequency-Wavenumber
  Spectra of Young Wind Waves}},}\ }\href {\doibase 10.1175/JPO-D-14-0237.1}
  {\bibfield  {journal} {\bibinfo  {journal} {J. Phys. Ocean.}\ }\textbf
  {\bibinfo {volume} {45}},\ \bibinfo {pages} {2484----2496} (\bibinfo {year}
  {2015})}\BibitemShut {NoStop}%
\bibitem [{\citenamefont {Hwang}\ \emph {et~al.}(2000)\citenamefont {Hwang},
  \citenamefont {Wang}, \citenamefont {Walsh}, \citenamefont {Krabill},\ and\
  \citenamefont {Swift}}]{Hwang}%
  \BibitemOpen
  \bibfield  {author} {\bibinfo {author} {\bibfnamefont {P.A.}\ \bibnamefont
  {Hwang}}, \bibinfo {author} {\bibfnamefont {D.W.}\ \bibnamefont {Wang}},
  \bibinfo {author} {\bibfnamefont {E.J.}\ \bibnamefont {Walsh}}, \bibinfo
  {author} {\bibfnamefont {W.B.}\ \bibnamefont {Krabill}}, \ and\ \bibinfo
  {author} {\bibfnamefont {R.N.}\ \bibnamefont {Swift}},\ }\bibfield  {title}
  {\enquote {\bibinfo {title} {{Airborne Measurements of the Wavenumber Spectra
  of Ocean Surface Waves. Part I: Spectral Slope and Dimensionless Spectral
  Coefficient}},}\ }\href
  {http://journals.ametsoc.org/doi/abs/10.1175/1520-0485(2001)031{\%}3C2753:AMOTWS{\%}3E2.0.CO{\%}3B2}
  {\bibfield  {journal} {\bibinfo  {journal} {J. Phys. Ocean.}\ }\textbf
  {\bibinfo {volume} {30}},\ \bibinfo {pages} {2753--2767} (\bibinfo {year}
  {2000})}\BibitemShut {NoStop}%
\bibitem [{\citenamefont {Romero}\ and\ \citenamefont
  {Melville}(2010)}]{Romero}%
  \BibitemOpen
  \bibfield  {author} {\bibinfo {author} {\bibfnamefont {L.}~\bibnamefont
  {Romero}}\ and\ \bibinfo {author} {\bibfnamefont {W~K}\ \bibnamefont
  {Melville}},\ }\bibfield  {title} {\enquote {\bibinfo {title} {{Airborne
  Observations of Fetch-Limited Waves in the Gulf of Tehuantepec}},}\
  }\href@noop {} {\bibfield  {journal} {\bibinfo  {journal} {Journal Of
  Physical Oceanography}\ }\textbf {\bibinfo {volume} {40}},\ \bibinfo {pages}
  {441--465} (\bibinfo {year} {2010})}\BibitemShut {NoStop}%
\bibitem [{\citenamefont {Melville}\ \emph {et~al.}(2016)\citenamefont
  {Melville}, \citenamefont {Lenain}, \citenamefont {Cayan}, \citenamefont
  {Kahru}, \citenamefont {Kleissl}, \citenamefont {Linden},\ and\ \citenamefont
  {Statom}}]{Melville}%
  \BibitemOpen
  \bibfield  {author} {\bibinfo {author} {\bibfnamefont {W.~K.}\ \bibnamefont
  {Melville}}, \bibinfo {author} {\bibfnamefont {L.}~\bibnamefont {Lenain}},
  \bibinfo {author} {\bibfnamefont {D.~R.}\ \bibnamefont {Cayan}}, \bibinfo
  {author} {\bibfnamefont {M.}~\bibnamefont {Kahru}}, \bibinfo {author}
  {\bibfnamefont {J.~P.}\ \bibnamefont {Kleissl}}, \bibinfo {author}
  {\bibfnamefont {P.~F.}\ \bibnamefont {Linden}}, \ and\ \bibinfo {author}
  {\bibfnamefont {N.~M.}\ \bibnamefont {Statom}},\ }\bibfield  {title}
  {\enquote {\bibinfo {title} {The modular aerial sensing system. journal of
  atmospheric and oceanic technology},}\ }\href@noop {} {\bibfield  {journal}
  {\bibinfo  {journal} {Journal of Atmospheric and Oceanic Technology}\
  }\textbf {\bibinfo {volume} {33}},\ \bibinfo {pages} {1169--1184} (\bibinfo
  {year} {2016})}\BibitemShut {NoStop}%
\bibitem [{\citenamefont {Lenain}\ and\ \citenamefont
  {Melville}(2017)}]{Lenain}%
  \BibitemOpen
  \bibfield  {author} {\bibinfo {author} {\bibfnamefont {L.}~\bibnamefont
  {Lenain}}\ and\ \bibinfo {author} {\bibfnamefont {W.~K.}\ \bibnamefont
  {Melville}},\ }\bibfield  {title} {\enquote {\bibinfo {title} {Measurements
  of the directional spectrum across the equilibrium saturation ranges of
  wind-generated surface waves.}}\ }\href@noop {} {\bibfield  {journal}
  {\bibinfo  {journal} {Journal of Physical Oceanography}\ }\textbf {\bibinfo
  {volume} {47}},\ \bibinfo {pages} {2123-- 2138} (\bibinfo {year}
  {2017})}\BibitemShut {NoStop}%
\bibitem [{\citenamefont {Denissenko}\ \emph {et~al.}(2007)\citenamefont
  {Denissenko}, \citenamefont {Lukaschuk},\ and\ \citenamefont
  {Nazarenko}}]{Denissenko2007}%
  \BibitemOpen
  \bibfield  {author} {\bibinfo {author} {\bibfnamefont {P.}~\bibnamefont
  {Denissenko}}, \bibinfo {author} {\bibfnamefont {S.}~\bibnamefont
  {Lukaschuk}}, \ and\ \bibinfo {author} {\bibfnamefont {S.}~\bibnamefont
  {Nazarenko}},\ }\bibfield  {title} {\enquote {\bibinfo {title} {Gravity wave
  turbulence in a laboratory flume},}\ }\href {\doibase
  10.1103/PhysRevLett.99.014501} {\bibfield  {journal} {\bibinfo  {journal}
  {Phys. Rev. Lett.}\ }\textbf {\bibinfo {volume} {99}},\ \bibinfo {pages}
  {014501} (\bibinfo {year} {2007})}\BibitemShut {NoStop}%
\bibitem [{\citenamefont {Nazarenko}\ and\ \citenamefont
  {Lukaschuk}(2016)}]{Nazarenko2016}%
  \BibitemOpen
  \bibfield  {author} {\bibinfo {author} {\bibfnamefont {S~V}\ \bibnamefont
  {Nazarenko}}\ and\ \bibinfo {author} {\bibfnamefont {S}~\bibnamefont
  {Lukaschuk}},\ }\bibfield  {title} {\enquote {\bibinfo {title} {{Wave
  Turbulence on Water Surface}},}\ }\href@noop {} {\bibfield  {journal}
  {\bibinfo  {journal} {Annual Review of Condensed Matter Physics}\ }\textbf
  {\bibinfo {volume} {7}},\ \bibinfo {pages} {61--88} (\bibinfo {year}
  {2016})}\BibitemShut {NoStop}%
\bibitem [{\citenamefont {Aubourg}\ \emph
  {et~al.}(2017{\natexlab{a}})\citenamefont {Aubourg}, \citenamefont {A.},
  \citenamefont {Peureux}, \citenamefont {Ardhuin}, \citenamefont {Sommeria},
  \citenamefont {Viboud},\ and\ \citenamefont {Mordant}}]{Aubourg2017}%
  \BibitemOpen
  \bibfield  {author} {\bibinfo {author} {\bibfnamefont {Q.}~\bibnamefont
  {Aubourg}}, \bibinfo {author} {\bibfnamefont {Campagne}\ \bibnamefont {A.}},
  \bibinfo {author} {\bibfnamefont {C.}~\bibnamefont {Peureux}}, \bibinfo
  {author} {\bibfnamefont {F.}~\bibnamefont {Ardhuin}}, \bibinfo {author}
  {\bibfnamefont {J.}~\bibnamefont {Sommeria}}, \bibinfo {author}
  {\bibfnamefont {S.}~\bibnamefont {Viboud}}, \ and\ \bibinfo {author}
  {\bibfnamefont {N.}~\bibnamefont {Mordant}},\ }\bibfield  {title} {\enquote
  {\bibinfo {title} {3-wave and 4-wave interactions in gravity wave
  turbulence},}\ }\href@noop {} {\bibfield  {journal} {\bibinfo  {journal}
  {Phys. Rev. Fluids}\ }\textbf {\bibinfo {volume} {2}},\ \bibinfo {pages}
  {114802} (\bibinfo {year} {2017}{\natexlab{a}})}\BibitemShut {NoStop}%
\bibitem [{\citenamefont {Onorato}\ \emph {et~al.}(2009)\citenamefont
  {Onorato}, \citenamefont {Cavaleri}, \citenamefont {Fouques}, \citenamefont
  {Gramstad}, \citenamefont {Janssen}, \citenamefont {Monbaliu}, \citenamefont
  {Osborne}, \citenamefont {Pakozdi}, \citenamefont {Serio}, \citenamefont
  {Stansberg}, \citenamefont {Toffoli},\ and\ \citenamefont
  {Trulsen}}]{Onorato2009}%
  \BibitemOpen
  \bibfield  {author} {\bibinfo {author} {\bibfnamefont {M.}~\bibnamefont
  {Onorato}}, \bibinfo {author} {\bibfnamefont {L.}~\bibnamefont {Cavaleri}},
  \bibinfo {author} {\bibfnamefont {S.}~\bibnamefont {Fouques}}, \bibinfo
  {author} {\bibfnamefont {O.}~\bibnamefont {Gramstad}}, \bibinfo {author}
  {\bibfnamefont {P.A.E.M.}\ \bibnamefont {Janssen}}, \bibinfo {author}
  {\bibfnamefont {J.}~\bibnamefont {Monbaliu}}, \bibinfo {author}
  {\bibfnamefont {A.~R.}\ \bibnamefont {Osborne}}, \bibinfo {author}
  {\bibfnamefont {C.}~\bibnamefont {Pakozdi}}, \bibinfo {author} {\bibfnamefont
  {M.}~\bibnamefont {Serio}}, \bibinfo {author} {\bibfnamefont {C.~T.}\
  \bibnamefont {Stansberg}}, \bibinfo {author} {\bibfnamefont {a.}~\bibnamefont
  {Toffoli}}, \ and\ \bibinfo {author} {\bibfnamefont {K.}~\bibnamefont
  {Trulsen}},\ }\bibfield  {title} {\enquote {\bibinfo {title} {{Statistical
  properties of mechanically generated surface gravity waves: a laboratory
  experiment in a three-dimensional wave basin}},}\ }\href {\doibase
  10.1017/S002211200900603X} {\bibfield  {journal} {\bibinfo  {journal} {J.
  Fluid Mech.}\ }\textbf {\bibinfo {volume} {627}},\ \bibinfo {pages} {235}
  (\bibinfo {year} {2009})}\BibitemShut {NoStop}%
\bibitem [{\citenamefont {Aubourg}(2016)}]{AubourgPhD}%
  \BibitemOpen
  \bibfield  {author} {\bibinfo {author} {\bibfnamefont {Q.}~\bibnamefont
  {Aubourg}},\ }\emph {\bibinfo {title} {Etude exp\'erimentale de la turbulence
  d'ondes \`a la surface d'un fluide. La th\'eorie de la Turbulence Faible \`a
  l'\'epreuve de la r\'ealit\'e pour les ondes de capillarit\'e et
  gravit\'e}},\ \href@noop {} {Ph.D. thesis},\ \bibinfo  {school} {Universit\'e
  Grenoble Alpes} (\bibinfo {year} {2016})\BibitemShut {NoStop}%
\bibitem [{\citenamefont {Phillips}(1958)}]{Phillips}%
  \BibitemOpen
  \bibfield  {author} {\bibinfo {author} {\bibfnamefont {O.M.}\ \bibnamefont
  {Phillips}},\ }\bibfield  {title} {\enquote {\bibinfo {title} {The
  equilibrium range in the spectrum of wind generated waves},}\ }\href@noop {}
  {\bibfield  {journal} {\bibinfo  {journal} {Journal Of Fluid Mechanics}\
  }\textbf {\bibinfo {volume} {4}},\ \bibinfo {pages} {426--434} (\bibinfo
  {year} {1958})}\BibitemShut {NoStop}%
\bibitem [{\citenamefont {Kuznetsov}(2004)}]{Kuz}%
  \BibitemOpen
  \bibfield  {author} {\bibinfo {author} {\bibfnamefont {E.A.}\ \bibnamefont
  {Kuznetsov}},\ }\bibfield  {title} {\enquote {\bibinfo {title} {Turbulence
  spectra generated by singularities},}\ }\href@noop {} {\bibfield  {journal}
  {\bibinfo  {journal} {JETP Lett.}\ }\textbf {\bibinfo {volume} {80}},\
  \bibinfo {pages} {83--89} (\bibinfo {year} {2004})}\BibitemShut {NoStop}%
\bibitem [{\citenamefont {Newell}\ \emph {et~al.}(2001)\citenamefont {Newell},
  \citenamefont {Nazarenko},\ and\ \citenamefont {Biven}}]{Biven}%
  \BibitemOpen
  \bibfield  {author} {\bibinfo {author} {\bibfnamefont {A.C.}\ \bibnamefont
  {Newell}}, \bibinfo {author} {\bibfnamefont {S.V.}\ \bibnamefont
  {Nazarenko}}, \ and\ \bibinfo {author} {\bibfnamefont {L.}~\bibnamefont
  {Biven}},\ }\bibfield  {title} {\enquote {\bibinfo {title} {{Wave turbulence
  and intermittency}},}\ }\href@noop {} {\bibfield  {journal} {\bibinfo
  {journal} {Physica D-Nonlinear Phenomena}\ }\textbf {\bibinfo {volume}
  {152}},\ \bibinfo {pages} {520--550} (\bibinfo {year} {2001})}\BibitemShut
  {NoStop}%
\bibitem [{\citenamefont {Aubourg}\ and\ \citenamefont
  {Mordant}(2015)}]{Aubourg2015}%
  \BibitemOpen
  \bibfield  {author} {\bibinfo {author} {\bibfnamefont {Quentin}\ \bibnamefont
  {Aubourg}}\ and\ \bibinfo {author} {\bibfnamefont {N.}~\bibnamefont
  {Mordant}},\ }\bibfield  {title} {\enquote {\bibinfo {title} {Nonlocal
  resonances in weak turbulence of gravity-capillary waves},}\ }\href {\doibase
  10.1103/PhysRevLett.114.144501} {\bibfield  {journal} {\bibinfo  {journal}
  {Phys. Rev. Lett.}\ }\textbf {\bibinfo {volume} {114}},\ \bibinfo {pages}
  {1--5} (\bibinfo {year} {2015})}\BibitemShut {NoStop}%
\bibitem [{\citenamefont {Nazarenko}(2013)}]{Nazarenko2013}%
  \BibitemOpen
  \bibfield  {author} {\bibinfo {author} {\bibfnamefont {Sergey}\ \bibnamefont
  {Nazarenko}},\ }\bibfield  {title} {\enquote {\bibinfo {title} {Sandpile
  behaviour in discrete water-wave turbulence},}\ }\href {\doibase
  10.1088/1742-5468/2006/02/L02002} {\bibfield  {journal} {\bibinfo  {journal}
  {J. Stat. Mech.: Theory and Experiment}\ }\textbf {\bibinfo {volume}
  {02002}},\ \bibinfo {pages} {1--8} (\bibinfo {year} {2013})},\ \Eprint
  {http://arxiv.org/abs/0510054} {0510054 [nlin]} \BibitemShut {NoStop}%
\bibitem [{\citenamefont {Aubourg}\ \emph
  {et~al.}(2017{\natexlab{b}})\citenamefont {Aubourg}, \citenamefont
  {Sommeria}, \citenamefont {Viboud},\ and\ \citenamefont
  {Mordant}}]{Aubourg2017b}%
  \BibitemOpen
  \bibfield  {author} {\bibinfo {author} {\bibfnamefont {Q.}~\bibnamefont
  {Aubourg}}, \bibinfo {author} {\bibfnamefont {J.}~\bibnamefont {Sommeria}},
  \bibinfo {author} {\bibfnamefont {S.}~\bibnamefont {Viboud}}, \ and\ \bibinfo
  {author} {\bibfnamefont {N.}~\bibnamefont {Mordant}},\ }\bibfield  {title}
  {\enquote {\bibinfo {title} {Combined stereoscopic wave mapping and particle
  image velocimetry},}\ }\href@noop {} {\bibfield  {journal} {\bibinfo
  {journal} {submitted to Exp. Fluids}\ } (\bibinfo {year}
  {2017}{\natexlab{b}})}\BibitemShut {NoStop}%
\end{thebibliography}%
\end{document}